\documentclass[%
 aip,
 amsmath,amssymb,
 reprint,apl%
]{revtex4-1}
\usepackage{xcolor}
\usepackage{graphicx}
\usepackage{dcolumn}
\usepackage{bm}

\usepackage[utf8]{inputenc}
\usepackage[T1]{fontenc}
\usepackage{mathptmx}
\usepackage{etoolbox}
\makeatletter
\def\@email#1#2{%
 \endgroup
 \patchcmd{\titleblock@produce}
  {\frontmatter@RRAPformat}
  {\frontmatter@RRAPformat{\produce@RRAP{*#1\href{mailto:#2}{#2}}}\frontmatter@RRAPformat}
  {}{}
}%
\makeatother
\begin{document}

\preprint{AIP/123-QED}

\title{Optimizing optical properties of bilayer PtSe$_2$: the role of twist angle and hydrostatic pressure}

\author{P. Jureczko}
\author{Z. Dendzik}
\author{M. Kurpas}
 \email{marcin.kurpas@us.edu.pl}
\affiliation{Institute of Physics, University of Silesia in Katowice, 41-500 Chorzów, Poland 
}%

\date{\today}

\begin{abstract}
Two-dimensional van der Waals materials offer exceptional tunability in their electronic properties. In this paper, we explore how twisting and hydrostatic pressure can be leveraged to engineer the electronic and optical characteristics of bilayer PtSe$_2$. Using state-of-the-art first-principles density functional methods, we calculate the electronic band structure and the imaginary part of the dielectric function across multiple twist angles and pressure values. We find, that at the twist angle $\theta=13.17^\circ$, bilayer PtSe$_2$, which is intrinsically an indirect semiconductor, transforms into a direct-gap semiconductor. Moreover, we demonstrate that hydrostatic out-of-plane pressure boosts near-infrared optical activity, further expanding the functional potential of PtSe$_2$ bilayers. 
The demonstrated high tunability of electronic and optical properties by twisting and pressure opens new application directions of PtSe$_2$ in optoelectronics.
\end{abstract}
\maketitle
\section{Introduction}
\label{Intro}
Two-dimensional (2D) transition metal dichalcogenides (TMDC) constitute a big family of materials intensively explored over the recent years.  
These, so called, \textit{post-graphene} materials display  plethora of properties which make them attractive for developing new paradigms of quantum technologies, while the strong structural stability  and compatibility with the available fabrication methods \cite{Meng2023,Xu2019} make them suitable for industrial and commercial usage. Particularly interesting in this context  are semiconductor TMDCs,  whose sizable intrinsic band gap allows for broad opto-electronic applications \cite{Koperski2015,Withers2015,Wagner2018,Lopez-Sanchez2013,Zeng2018}. 
Additionally, spin-valley locking in non-centrosymmetric semiconductor TMDCs, enabled by a large spin-orbit coupling in valence bands \cite{Zhu_2011} allows for optical valley control and selective spin population, creating great conditions for optospintronics and valleytronics applications \cite{Mak2014,Luo2017,Langer2018,Abdukayumov2024,Mueller2018,Dutta2023}.

Monolayer PtSe$_2$ is TMDC characterized by robust environmental stability and high  room temperature carrier mobility \cite{Zhang2014,Zhao2017,Yang2019,Bonell_2022}. Like its bulk counterpart, monolayer PtSe$_2$ crystallizes in the centrosymmetric octahedral 1T phase. The 1T phase often coexists with the trigonal prismatic 1H phase, which is a metastable state of higher energy, and under light irradiation, it evolves to the 1T phase \cite{Xu2021}.  
In the bulk, PtSe$_2$ is a semimetal with a small overlap of the valence and conduction bands \cite{Guo_1986}.
In the monolayer limit, it  is a semiconductor with a sizable indirect band gap enclosed in a range of 1.2-2\,eV.~\cite{ptse2_2015_wang,Yu2018}

Owing to the wide energy range of photon transitions, from ultraviolet to mid-infrared \cite{Sajjad2018,pribusova_2021,Bae2021}, PtSe$_2$ is a promising material for building optoelectronic devices, such as sensors \cite{Wagner2018}, photodetectors  \cite{Zeng2018,Yang2019} or photodiodes \cite{Yim2018}. 
Due to large refracting index it exhibits good lensing properties useful for building ultra-thin flat lenses achieving diffraction-limited focusing and imaging \cite{Lin2020}.    
The spectrum of applicability  in electronic and  optoelectronic devices can be further extended by applying strain \cite{Li_JMCC_2016,Li2016,Deng2018,Amairi_2025} or by exploiting the thickness dependence of the electronic and optical properties \cite{Yu2018,Ciarrocchi2018,multilayers_PtSe2_Villaos2019,structural_parameters_Li2021,Zheng2022}, which gives additional control over the device parameters. 
Additionally, in multilayered samples, stacking acts as en extra knob for tailoring materials properties, for instance, photon transition rates \cite{Fang2019}.
However, the significant momentum mismatch between the valence and conduction band edges precludes PtSe$_2$ from   applications as efficient light emitting devices. %

Very recently, another method of tailoring materials properties by twisting \cite{Carr2017} has become a powerful approach for tuning the properties of two-dimensional materials. At its core, it leverages quantum interactions between atoms in stacked layers, which change when the layers are rotated relative to one another. 
This rotation generates a Moir\'{e} pattern, profoundly influencing the material’s electronic characteristics.
Several fascinating phenomena have been demonstrated to arise due to twist-angle induced correlations, including strongly correlated states \cite{Li2010,Morell2010,Dean2013,Carr2017},  superconductivity \cite{Cao2018,Klebl2022}, or electronic decoupling due to large momentum mismatch \cite{Rickhaus2020}. 
Twist-induced structural changes map also on the  optical properties resulting in the observation of long living Moir\'{e} excitons \cite{Wu2017,Tran2019,He2021,Huang2022} and indirect to direct band gap transitions in semiconductors \cite{Gao2020,Xin2022,Gupta2024}.
In this context, PtSe$_2$ is very promising material due to PtSe$_2$ a small energy barrier for the kinetic evolution of interlayer structures, allowing  dynamical control of over stackings and twist angles \cite{Xu2021_2}.

In this paper, we study the structural, electronic and optical properties of twisted bilayer PtSe$_2$. Using first-principles calculations, we examine the electronic band structure and optical properties by means of the imaginary part of the dielectric function for five different twist angles ranging from $0^{\circ}$ to $60^{\circ}$. 
We show that by a fine tuning the twist angle $\theta$, the optical activity of bilayer PtSe$_2$ can be widely controlled. Specifically, for $\theta=13.17^\circ$, the electronic band gap becomes direct, enabling  PtSe$_2$ as a luminescent material. 
We also show, that optical activity of PtSe$_2$ bilayer in the near infrared spectrum can be greatly enhanced by applying hydrostatic pressure. 
This method has been recently considered as control knob for the spin-orbit proximity effect \cite{Fulop2021,Kedves2023,MM2024}  or spin relaxation anisotropy \cite{Jureczko_2025}.

Our findings demonstrate that the optical properties of bilayer PtSe$_2$ can be precisely and extensively tuned using currently available experimental techniques. 
This paves the way for  new applications of PtSe$_2$, particularly in the development of thin optoelectronic devices, unlocking new possibilities for next-generation technology.

The paper is organized as follows: In the next section, we present the research methodology, including details of the numerical calculations. In Section II, we outline the main results of the paper and discuss their interpretation. Finally, Section III concludes the paper with a summary of the findings.

\section{Methods}\label{sec:methods}
First-principles calculations were done using the Q{\sc{uantum}} ESPRESSO open-source DFT package \cite{Giannozzi_2009,Giannozzi_2017}. We used the Optimized Norm-Conserving Vanderbilt  SG15  pseudopotentials with the Perdew-Burke-Ernzerhof\,(PBE) version of the generalized gradient approximation\,(GGA) exchange-correlation functional \cite{hamnann_oncv, PBE}.  
Self-consistency was achieved with the kinetic energy cutoff 50\,Ry for the wave function and 200\,Ry for charge density. To exclude the interaction between copies of PtSe$_2$ layers, we introduced a vacuum of 17\,\AA\, in the direction perpendicular to the two-dimensional film. Wan der Waals interaction was taken into accouunt using the semiempirical Grimme DFT-D3 dispersion correction. For band gap correction, we used the hybrid Heyd-Scuseria-Ernzerhof\,(HSE) functional, assuming the typical  20\%  contribution of the Fock exchange \cite{HSE}.

Starting from the experimental value a = 3.7\,\AA \cite{ptse2_2015_wang} the lattice constant for monolayer PtSe$_2$ was optimized using variable cell relaxation based on quasi-Newton scheme implemented in  Q{\sc{uantum}} ESPRESSO. We found a = 3.748\,\AA\, for monolayer and AA bilayer, while for $\theta$ = 60$^{\circ}$, the lattice constant adopts a = 3.727\,\AA. Systems with twist angles $\theta$ = 13.27$^\circ$, 21.79$^\circ$ and 32.20$^\circ$ were constructed using the scheme proposed in Ref. \cite{Uchida2014}. Optimized structure parameters are presented in  Table \ref{tab:table_1} and in Fig. S1 in the Supplemental Material.
Self-consistency was achieved with a 21x21x1 Monkhorst-Pack grid for monolayer, AA and $\theta$ = 60$^\circ$ bilayer. For twist angles $\theta$ = 13.27$^\circ$, 21.79$^\circ$ and 32.20$^\circ$ the k-points mesh was 6x6x1, 9x9x1 and 6x6x1, respectively. 

Investigating the optical properties of the materials is possible by analyzing the linear response of the electronic system to an electromagnetic perturbation, such as incoming light. The material response is determined by dielectric function $\varepsilon(\omega) = \varepsilon_{1}(\omega)+i\varepsilon_{2}(\omega)$, composed of real $\varepsilon_{1}(\omega)$ and imaginary $\varepsilon_{2}(\omega)$ part. The latter corresponds to the energy absorption of the system and the transition between occupied and unoccupied states. Using Random Phase Approximation, the imaginary complex dielectric tensor $\varepsilon_{2}(\omega)$ can be define in the following form \cite{Ambrosch-Draxl2006,Adler1962}:  
\begin{equation*}
\begin{split}
    Im\, \varepsilon_{ij}(\omega)  = \frac{\hbar^{2}e^{2}}{\pi m^{2}\omega^{2}}\sum_{n,n'} \int_{BZ} \langle n'\textbf{k} |\hat{p}_{i}| n\textbf{k} \rangle \langle n'\textbf{k} |\hat{p}_{j}| n\textbf{k} \rangle \\ \Bigl ( f_{0}(\varepsilon_{n,\textbf{k}}) - f_{0}(\varepsilon_{n',\textbf{k}}) \Bigl) \delta(\varepsilon_{n',\textbf{k}} - \varepsilon_{n,\textbf{k}} - \omega)\,d\textbf{k}
\end{split}
\end{equation*}
where $n,n'$ belong to the valence and conduction band, $f_{0}$ is a Fermi Dirac distribution, $\hat{p}_{i,j}$ describe the momentum matrix elements between bands $n$ and $n'$ and crystal momentum $\textbf{k}$. The real part of the dielectric function can be obtained from the Kramers-Kronig transformation. 
We consider only interband transitions occurring between different bands. Such an approach is valid for semiconductors. The dielectric tensor consists of six components related to the system symmetry. For a hexagonal structure, the imaginary dielectric tensor is diagonal with two independent components, $\varepsilon_{xx}$ = $\varepsilon_{yy} \neq \varepsilon_{zz}$. We calculated the imaginary part of the dielectric function for PtSe$_2$ monolayer and twisted bilayers using the post-processing code $\textit{epsilon.x}$ implemented in Q{\sc{uantum}} ESPRESSO. The mesh convergence for non-self-consistent calculations was achieved with 30x30x1 k-points for monolayer, $\theta$ = 0$^\circ$ and 60$^\circ$ bilayers. For angles 13.27$^\circ$, 21.79$^\circ$ and 32.20$^\circ$ the mesh was 9x9x1, 30x30x1 and 15x15x1. We use the broadening parameter $\textit{intersmear}$ for the interband contribution with a value of 0.1\,eV in optical transition calculations.
\section{Results and Discussion}
\subsection{Electronic structure}
We begin with discussing the electronic properties of monolayer PtSe$_2$ which constitutes  the benchmark material for bilayer systems.  In Fig. \ref{fig_bnd_1} b) we show the calculated band structure of monolayer PtSe$_2$. As can be seen, single layer PtSe$_2$ is an indirect gap semiconductor with  the gap   1.2\, eV obtained for the PBE exchange-correlation potential. Each band is two-fold degenerate owing to inversion symmetric crystal structure of the monolayer. In connection with strong spin-orbit coupling it leads to fast spin relaxation \cite{Kurpas2021}. 
The lowest two conduction bands (four including spin) form a distinct closed manifold, separated from higher-lying bands by 1.5 eV (not shown). This unique band structure topology fundamentally shapes the low-energy optical spectrum, restricting transitions from the valence band exclusively to these two conduction bands

\begin{figure}
    \centering
    \includegraphics[width = 0.99\columnwidth]{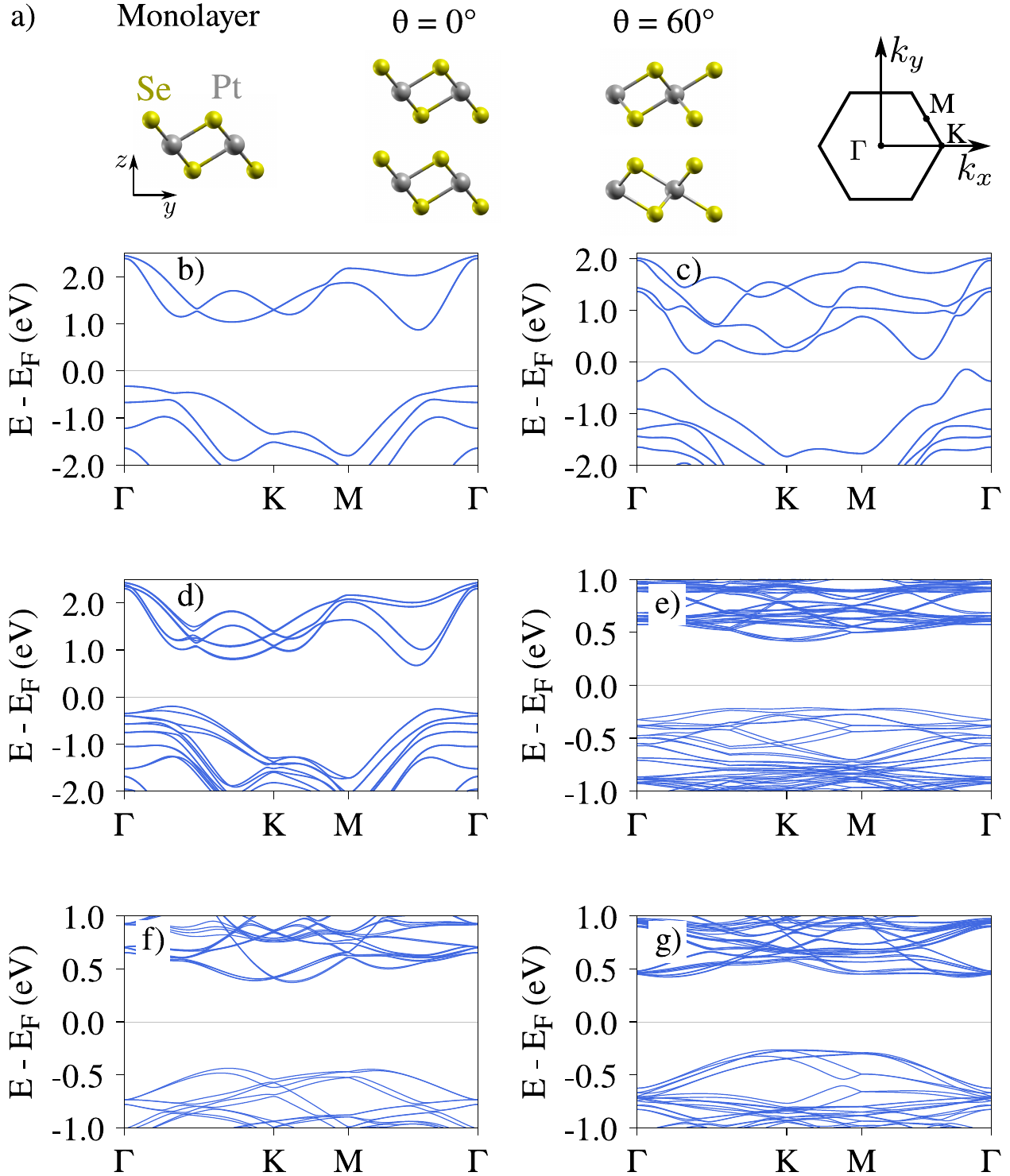}
    \caption{Electronic properties of mono and bilayer PtSe$_2$:  (a) Sketch of monolayer, 0$^\circ$ and 60$^\circ$ twisted bilayer PtSe$_2$ crystal structures and the corresponding first Brillouin zone with marked high symmetry points. Calculated relativistic bands structures  for single layer (b), and twisted bilayers are shown in (c)- (g) for 0$^\circ$, 60$^\circ$, 13.17$^\circ$, 21.79$^\circ$ and 32.20$^\circ$ respectively. \label{fig_bnd_1} }
\end{figure}
Let us now discuss bilayer PtSe$_2$. We prepare bilayer samples by stacking two identical centrosymmetric 1T phase layers one above another and apply a twist of the angle $\theta$. 
Due to the $C_3$ rotation axis in a single layer the two limiting twist angles for PtSe$_2$ bilayers are $\theta = 0^\circ$,  and $\theta = 60^\circ$. 
The former ($\theta = 0^\circ$) corresponds to the AA stacking of the monolayers, see Fig. \ref{fig_bnd_1} a).  Like in bulk and single layer PtSe$_2$ the structure of AA stacked bilayer has $D_{3d}$ point group symmetry. The $D_{3d}$ point group contains the inversion operation, which guarantees spin degeneracy of bands at any k-point. 
Also like the monolayer,  the AA bilayer is an indirect gap semiconductor, but with a much smaller band gap  0.2\,eVm see Fig. \ref{fig_bnd_1} e) and Table \ref{tab:table_1}.
We attribute the gap reduction to the attracting interlayer interaction manifested by a considerable charge accumulation in the interlayer region. This is confirmed by a large peak in the charge density profile $\Delta \rho (z)$, see Fig. S2 a) in the Supplemental Material.

A qualitatively different physics we observe for the twist angle $\theta = 60^\circ$ (AB stacking).  In this case, the crystal structure is equivalent to the AA case but with one layer twisted by $180^\circ$, see Fig. \ref{fig_bnd_1} a), and respects $D_{3h}$ point group symmetry.  
Now, the bottom Se atom from the upper layer sits on top of another Se atom from the lower layer, which energetically is an unfavorable state and costs an extra 180\,meV/atom compared to the AA case. 
The binding energy for the AB configuration is twice smaller than for the AA case, consistent with other reports \cite{Fang2019}, but is still negative, see Table \ref{tab:table_1}, suggesting  the structural stability of the AB stacking. 
Some authors report softening of acoustic modes in phonon spectrum close to the $\Gamma$. \cite{Fang2019} However, it is an artifact attributed  to numerical precision rather than to the structural stability issue. \cite{Fang2019,Kandemir2018} 
Unlike the AA case, now  $\Delta \rho(z)$ shows charge depletion in the interlayer region and $\Delta\approx 0$ at z = 0, Fig. S2 a) in the Supplemental Material, which indicates a weaker electronic interaction between the layers than for AA case, due to the increased interlayer distance. 
Such interlayer interaction weakening is linked to low stability local stackings present in twisted systems \cite{Liu2014_vdw,Nayak2017,Yan2019,Ahn2024}.

In terms of electronic properties, $\theta = 60^\circ$ case is very similar to monolayer PtSe$_2$. 
The band structure remains that of the monolayer PtSe$_2$, Fig.~\ref{fig_bnd_1} b) and d) ) with the number of bands doubled and slightly shifted due to different crystal potentials for the top and bottom layers. Unlike the monolayer, whose Bloch states are spin degenerate due to space inversion symmetry, the states of $D_{3h}$ symmetric AB bilayer are spin split. However, spin degeneracy is maintained along the $\Gamma$M direction in the Brillouine zone, which lies on the mirror symmetry plane.   

Compared to the monolayer, the bandgap of $60^\circ$ twisted bilayer reduces by 0.35\,eV to the value 0.85 \,eV in the case of PBE potential, and by 0.29\,eV to the value 1.9\,eV for the hybrid HSE pseudopotential, see Table \ref{tab:table_1}.

For the intermediate twist angles $\theta = 13.17^\circ$, $\theta = 21.79^\circ$ and $\theta = 32.20^\circ$ the structural properties of PtSe$_2$ bilayers are  similar the each other. They are characterized by similar binding energy, interlayer distance  and charge density profile.

The case $\theta = 13.17^\circ$ is the most interesting.  In contrast to $\theta = 0^\circ$ and $\theta = 60^\circ$ for this twist angle the band gap is almost direct, see Fig. \ref{fig_bnd_1} a). The conduction band edge is only  slightly off the valence band edge, by $\delta_k = 0.015$\,\AA$^{-1}$, while the energy difference between direct and indirect gap is  only 2.3\,meV. For such minor discrepancies, the band gap can effectively be treated as direct, highlighting how twisting can transform PtSe$_2$ into a luminescent material.

It is known that bare exchange correlation potentials systematically underestimate the band gap of semiconductors due to self-interaction errors and missing long-range exact exchange.
More accurate approaches, such as GW or hybrid functionals incorporating a fraction of exact Hartree-Fock exchange, yield significantly more realistic band structures but come at a considerably higher computational cost. An extensive analysis of the band structures and optical properties of PtSe$_2$ systems shows that hybrid functional and GW results can be well reproduced by applying a scissor shift to the conduction bands of PBE results \cite{Tharrault2025}. Thus, we recalculate the band structures using the hybrid HSE potential for configurations with reasonable number of atoms. Our results, shown in Fig. S6 in the Supplemental Material, are in line with Ref. \cite{Tharrault2025} and indicate that the band structures obtained using the hybrid HSE functional can be well reproduced by the bare PBE functional corrected by a constant scissor shift of conduction bands.

\begin{table}
\caption{\label{tab:table_1}Structural and electronic properties of monolayer and twisted bilayer PtSe$_2$ from first principles.}
\begin{tabular}{cccccc}
\hline
\hline
$\theta$\,($^\circ$) & a\,(\AA) &d$_{\rm{inter}}$\,(\AA) & E$_{g}^{\rm{PBE}}$\,(eV) &E$_{g}^{\rm{HSE}}$\,(eV) \\ \hline\hline 
 ML & 3.748 & - & 1.2 & 2.19  \\ 
0 & 3.748 &2.373 & 0.2 & 1.08  \\ 
60 &3.727 &   3.742 & 0.85 &   1.9 \\ 
13.17 & 16.34 & 2.836 & 0.63  & - \\ 
21.79 & 9.92 &3.376 & 0.81  & - \\ 
32.20 &  14.67 & 3.126 &  0.69 & -\\ 
\end{tabular}
\end{table}

\subsection{Optical properties}
Recent experiments supported by theoretical analysis suggest that optical absorption in PtSe$_2$ in the near-infrared to visible range is dominated by single electron direct transitions, while phonon assisted transitions are negligible \cite{Tharrault2025}.  Therefore  it is justified to  use the single electron approach and describe optical properties 
by means of  the dielectric function $\varepsilon(\omega)$: $\varepsilon(\omega) = \varepsilon_{1}(\omega)+i\varepsilon_{2}(\omega)$, composed of real $\varepsilon_{1}(\omega)$ and imaginary $\varepsilon_{2}(\omega)$ part. The latter corresponds to the energy absorption due to transitions from occupied to the unoccupied states.

\begin{figure}[h]
    \centering
    \includegraphics[width = 0.99\columnwidth]{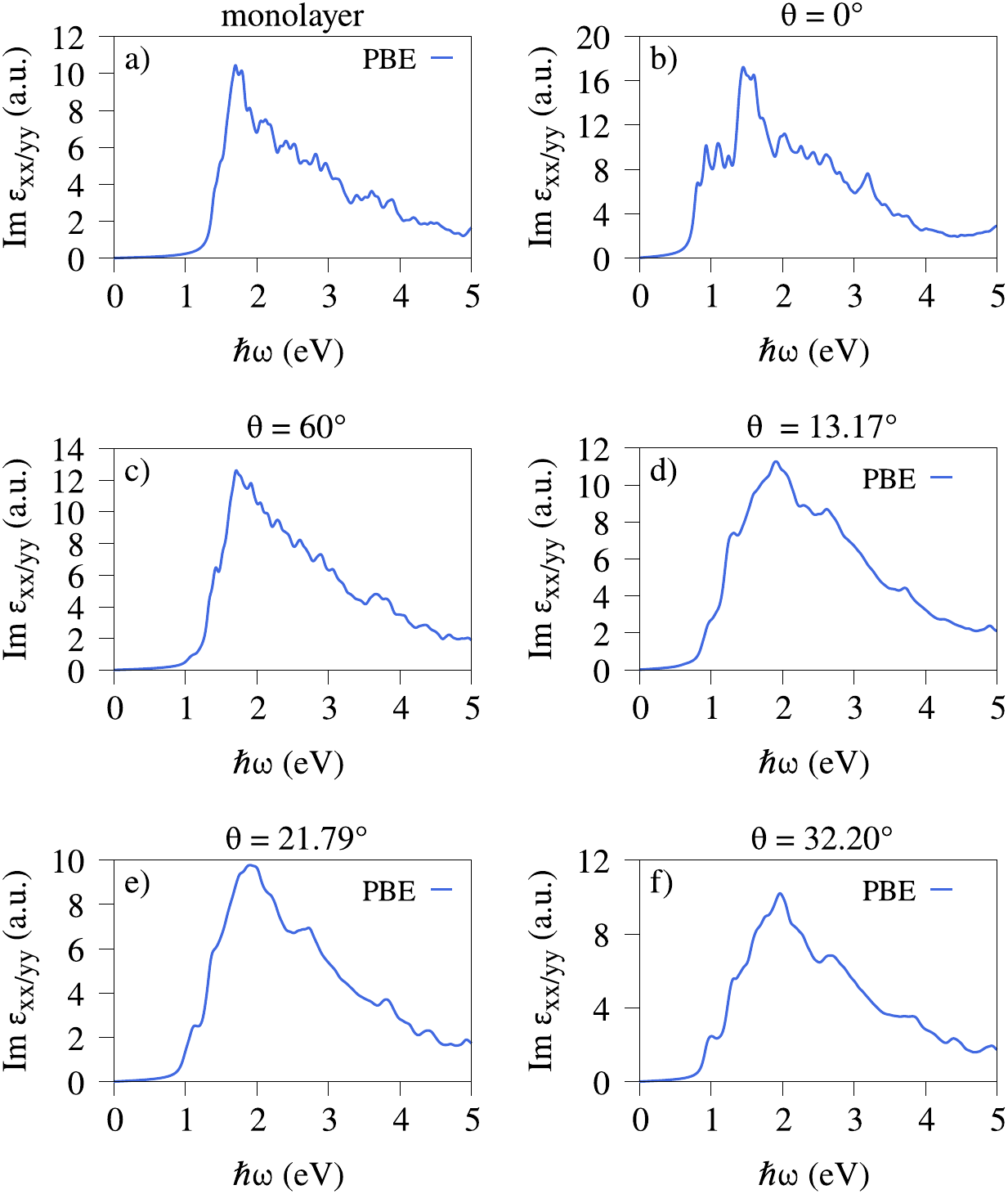}
    \caption{Imaginary part of dielectric fucntion for monolayer (a), and twisted bilayer PtSe$_2$ (b)-(g).\label{fig:opt}}
\end{figure}

In Fig. \ref{fig:opt}  we show the calculated imaginary part of dielectric tensor $\epsilon_{xx/yy}$.  
For all cases, also for the monolayer, the biggest optical activity is found for photon energies between 1.8\,eV - 2\,eV. 
At these energies, transitions from occupied valence states occur exclusively to the lowest conduction band manifold, which is separated from higher-lying bands by approximately 1.5 eV. This energy gap between conduction band manifolds leads to a reduction in the dielectric function within the 3\,eV to 5\,eV range, as illustrated in Fig. S3 of the Supplemental Material.

For the $\theta=0^\circ$ bilayer, several peaks appear between 1\,eV and 2\,eV, resulting from its significantly smaller band gap compared to both the monolayer and the $\theta=60^\circ$ bilayer. In cases with intermediate twist angles, see Fig. \ref{fig:opt} d)-f),  the imaginary part of the dielectric function resembles that of the $\theta=60^\circ$ bilayer but with a slightly broader peak. Overall, the most pronounced effect on the dielectric function is observed for the $\theta=60^\circ$ bilayer due to its considerably reduced band gap.

\begin{figure}
    \centering
    \includegraphics[width=0.99\columnwidth]{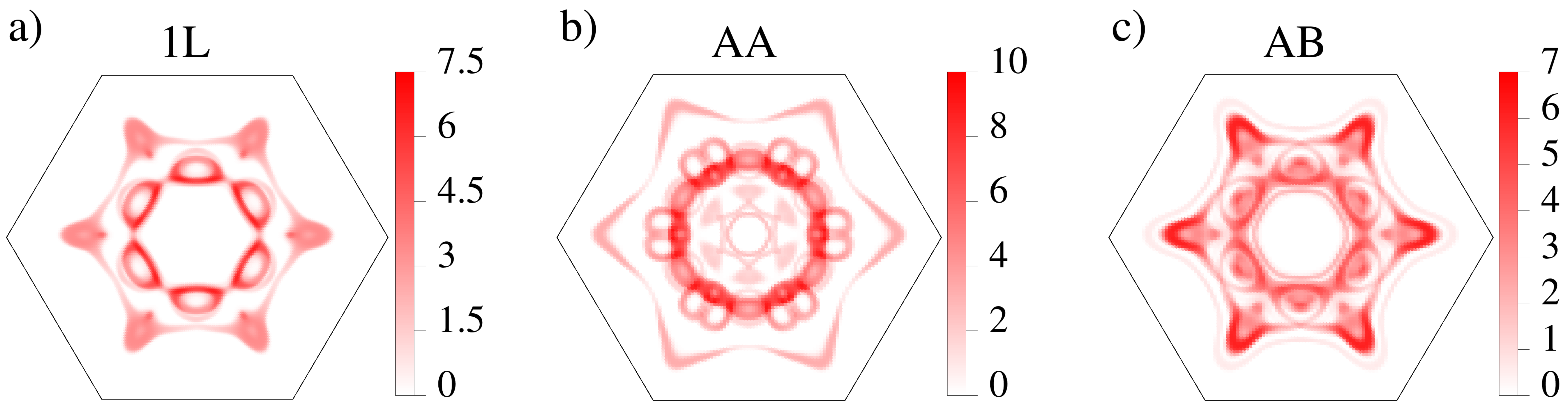}
    \caption{Momentum matrix elements between valence and conduction bands corresponding to main absorption peak at $\hbar \omega \sim 2$\,eV in the energy vicinity $\pm 0.7$\,eV around the peak center.  Here a non-relativistic approximation is used, see Fig. S3 a)-c). }
    \label{fig:mtx_bz}
\end{figure}

To further investigate the origin of the main peak in the dielectric function, we calculated the momentum (dipole) matrix elements between valence and conduction bands within a small energy range of approximately $\pm 0.7$\,eV around the peak center at $\hbar \omega \approx 2$\,eV. Since spin-orbit interaction has only a minor effect on the overall absorption spectrum (see Figs S3-S6 in the Supplemental Material), we focus on the non-relativistic case and consider the monolayer, $\theta=0^\circ$, and $\theta=60^\circ$ bilayers to optimize computational efficiency. The results, shown in Fig. \ref{fig:mtx_bz}, reveal a strong resemblance between the monolayer and $\theta=60^\circ$ bilayer, reflecting their similar band structures. 
However, a closer examination uncovers notable differences: for the monolayer (Fig. \ref{fig:mtx_bz} a)), the highest transition probability occurs at the inner contour, whereas in the AB-stacked bilayer (Fig. \ref{fig:mtx_bz} c)), the outer contour exhibits a higher transition probability.

In the AA ($\theta=0^\circ$) case, the most active Brillouin zone region is the hexagon formed around the $\Gamma$ point, see Fig. \ref{fig:mtx_bz} b), in particular the points in the $\Gamma$K path. 

\subsection{Effects of hydrostatic pressure}
\begin{figure}[h]
    \centering
    \includegraphics[width = 0.8\columnwidth]{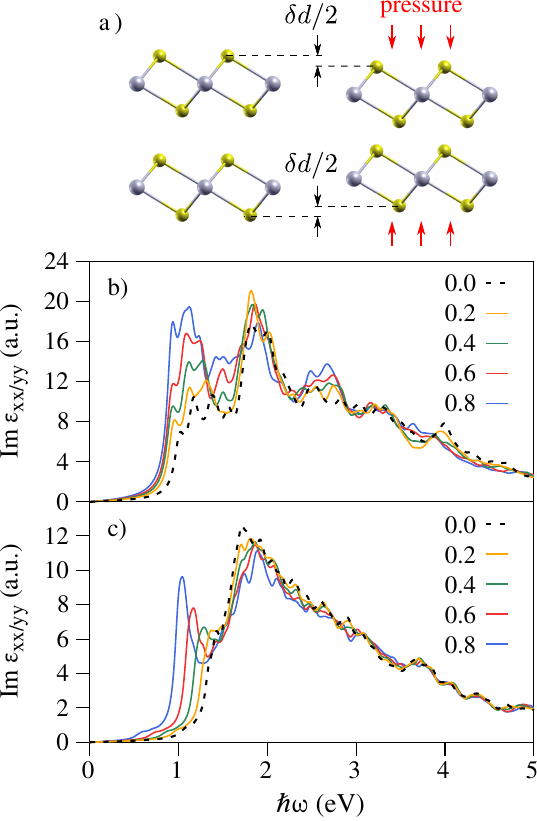}
    \caption{Bilayer PtSe$_2$ under hydrostatic pressure: (a) a sketch of simulating hydrostatic pressure in first-principles calculations. Starting from the equilibrium geometry, the outer Se atoms are moved towards the center of the heterostructure. Next, keeping their $z$-coordinates fixed the structure is relaxed. (b) Imaginary part of the dielectric function versus photon energy for the AA  bilayer ($\theta=0^\circ$) and for various values of structure thickness compression $\delta$d given in angstrom. (b) same as (a) but  for the AB stacking ($\theta=60^\circ$). \label{fig:strain_opt} }
\end{figure}

We now focus on the possibility of tuning the optical properties of PtSe$_2$ bilayer through the application of hydrostatic pressure.
In our first-principles calculations, we simulate hydrostatic pressure in the \textit{z} (out-of-plane) direction by reducing the bilayer thickness by $\delta \rm{d}$, ensuring that the $z$ coordinates of the outer Se atoms remain fixed at $z \pm \delta \rm{d}/2$ (see Fig. \ref{fig:strain_opt} a)). Next, we allow the structure to relax internal forces and adjust to the in-plane pressure by performing variable cell relaxation while keeping the $z$ coordinates of the outer Se atoms fixed.

The resulting structural parameters, energy gap, and pressure in the $z$ direction are presented in Fig. S1 in the Supplemental Material. As expected, vertical pressure reduces the interlayer distance while inducing a slight expansion of the lattice constant and a flattening of the PtSe$ _2$ layers. Within the experimentally accessible pressure range of 0–3 GPa \cite{Fulop2021}, controlled stepwise by $\delta \rm{d} =  \lbrace$ 0.2\,\AA, 0.4\,\AA, 0.6\,\AA, 0.8\,\AA, 1\,\AA$\rbrace$, the lattice constant increases by 3\% and 1.9\%, while the interlayer distance $\delta \rm{d}$ decreases by 21\% and 17\% for the AA and AB stacking configurations, respectively. Additionally, the thickness of the PtSe$_2$ layer grows with applied pressure, reaching a maximum increase of 6.4\% in the AB-stacked bilayer and 9\% in the AA-stacked bilayer for $\delta \rm{d} = 1$\,\AA.

In Fig. \ref{fig:strain_opt} b) and Fig. \ref{fig:strain_opt} c), we show the variation of the imaginary part of dielectric function under applied pressure for AA and AB PtSe$_2$ bilayers, respectively. In both cases, a significant enhancement in optical activity is observed at photon energies between 0.8\,eV and 1.5\,eV. However, the nature of this enhancement differs between the two stacking configurations. 
In the AA-stacked bilayer, we observe an increase in the amplitudes of peaks already present at zero pressure, accompanied by a slight frequency shift toward lower photon energies. By analyzing the dipole matrix elements, we find that this increase in optical activity is directly linked to the energy downshift of the electronic pocket located between the $\Gamma$ and K points, as shown in Fig. S5 in the Supplemental Material.

In the AB-stacked bilayer, the optical response to applied pressure is significantly stronger than in the AA-stacked case. At zero strain, no active optical transitions occur at $\hbar \omega \sim 1$\,eV. However, for $\delta > 0.2$\,\AA, two electronic pockets, located between the $\Gamma$M and $\Gamma$K high symmetry points, begin to contribute, with their amplitude increasing as the Fermi contour expands (see Fig. S5 in the Supplemental Material). The overall frequency downshift in the AB-stacked bilayer is approximately 0.75 eV, driven by band gap reduction, and it substantially exceeds the shift observed in the AA-stacked case.
The observed effects of hydrostatic pressure on optical properties are similar to those induced by biaxial in-plane strain \cite{Deng2018}. These similarities can be attributed to the stretching of PtSe$_2$ layers that accompanies the compression of the bilayer.

Due to computational constraints, we did not consider intermediate twist angles in detail. However, an example  result obtained for $\theta = 21.79^\circ$ shown in Fig. S7 of the Supplemental Material indicates similar but  slightly weaker response of  
$\epsilon_{xx/yy}$ on the applied strain as for the limiting twist angles $\theta=0^\circ$ and $\theta=60^\circ$. 
For the same reason, we calculated $\epsilon_{xx/yy}$ using hybrid exchange-correlation functionals only for the smallest systems: monolayer, AA and AB bilayers. Results are shown in Fig. S6 in the Supplemental Material. As already mentioned above, in the energy range considered here, the PBE results remain conclusive, taking into account a constant energy shift between PBE and the HSE results. 
\section{Conclusions}
We have investigated the electronic and optical properties of monolayer and twisted  bilayer PtSe$_2$. Using first principles calculations we have shown, that for the twist angle 13.17$^\circ$ the electronic band structure of the bilayer transforms into a direct gap semiconductor, opening new perspectives for applications of PtSe$_2$ as a light emitting devices. 
Furthermore, we have systematically analyzed the impact of hydrostatic pressure on optical absorption  driven in PtSe$_2$ by direct single electron transitions \cite{Tharrault2025}. 
Our findings reveal a remarkable tunability of optical properties under pressure, particularly in the near-infrared wavelength range, opening up potential avenues for  PtSe$_2$  in optoelectronics.

\section*{Acknowledgment}
P. J. and M.K.~acknowledge  financial support provided by the National Center for Research and Development (NCBR) under the V4-Japan project BGapEng V4-JAPAN/2/46/BGapEng/2022 and support from the Interdisciplinary Centre for Mathematical and Computational Modelling (ICM), University of Warsaw (UW), within grant no. G83-27. 

\section*{References}
\bibliography{sample}

\begin{thebibliography}{70}%
\makeatletter
\providecommand \@ifxundefined [1]{%
 \@ifx{#1\undefined}
}%
\providecommand \@ifnum [1]{%
 \ifnum #1\expandafter \@firstoftwo
 \else \expandafter \@secondoftwo
 \fi
}%
\providecommand \@ifx [1]{%
 \ifx #1\expandafter \@firstoftwo
 \else \expandafter \@secondoftwo
 \fi
}%
\providecommand \natexlab [1]{#1}%
\providecommand \enquote  [1]{``#1''}%
\providecommand \bibnamefont  [1]{#1}%
\providecommand \bibfnamefont [1]{#1}%
\providecommand \citenamefont [1]{#1}%
\providecommand \href@noop [0]{\@secondoftwo}%
\providecommand \href [0]{\begingroup \@sanitize@url \@href}%
\providecommand \@href[1]{\@@startlink{#1}\@@href}%
\providecommand \@@href[1]{\endgroup#1\@@endlink}%
\providecommand \@sanitize@url [0]{\catcode `\\12\catcode `\$12\catcode `\&12\catcode `\#12\catcode `\^12\catcode `\_12\catcode `\%12\relax}%
\providecommand \@@startlink[1]{}%
\providecommand \@@endlink[0]{}%
\providecommand \url  [0]{\begingroup\@sanitize@url \@url }%
\providecommand \@url [1]{\endgroup\@href {#1}{\urlprefix }}%
\providecommand \urlprefix  [0]{URL }%
\providecommand \Eprint [0]{\href }%
\providecommand \doibase [0]{http://dx.doi.org/}%
\providecommand \selectlanguage [0]{\@gobble}%
\providecommand \bibinfo  [0]{\@secondoftwo}%
\providecommand \bibfield  [0]{\@secondoftwo}%
\providecommand \translation [1]{[#1]}%
\providecommand \BibitemOpen [0]{}%
\providecommand \bibitemStop [0]{}%
\providecommand \bibitemNoStop [0]{.\EOS\space}%
\providecommand \EOS [0]{\spacefactor3000\relax}%
\providecommand \BibitemShut  [1]{\csname bibitem#1\endcsname}%
\let\auto@bib@innerbib\@empty
\bibitem [{\citenamefont {Meng}\ \emph {et~al.}(2024)\citenamefont {Meng}, \citenamefont {Yang}, \citenamefont {Dai}, \citenamefont {Tang}, \citenamefont {He}, \citenamefont {Gu}, \citenamefont {Jiang}, \citenamefont {Ding},\ and\ \citenamefont {Xu}}]{Meng2023}%
  \BibitemOpen
  \bibfield  {author} {\bibinfo {author} {\bibfnamefont {S.}~\bibnamefont {Meng}}, \bibinfo {author} {\bibfnamefont {Y.}~\bibnamefont {Yang}}, \bibinfo {author} {\bibfnamefont {X.}~\bibnamefont {Dai}}, \bibinfo {author} {\bibfnamefont {Y.}~\bibnamefont {Tang}}, \bibinfo {author} {\bibfnamefont {M.}~\bibnamefont {He}}, \bibinfo {author} {\bibfnamefont {Y.}~\bibnamefont {Gu}}, \bibinfo {author} {\bibfnamefont {R.}~\bibnamefont {Jiang}}, \bibinfo {author} {\bibfnamefont {F.}~\bibnamefont {Ding}}, \ and\ \bibinfo {author} {\bibfnamefont {H.}~\bibnamefont {Xu}},\ }\href {\doibase https://doi.org/10.1002/adfm.202312165} {\bibfield  {journal} {\bibinfo  {journal} {Advanced Functional Materials}\ }\textbf {\bibinfo {volume} {34}},\ \bibinfo {pages} {2312165} (\bibinfo {year} {2024})},\ \Eprint {http://arxiv.org/abs/https://advanced.onlinelibrary.wiley.com/doi/pdf/10.1002/adfm.202312165} {https://advanced.onlinelibrary.wiley.com/doi/pdf/10.1002/adfm.202312165} \BibitemShut {NoStop}%
\bibitem [{\citenamefont {Xu}\ \emph {et~al.}(2019)\citenamefont {Xu}, \citenamefont {Zhang}, \citenamefont {Liu}, \citenamefont {Zhang}, \citenamefont {Sun}, \citenamefont {Guo}, \citenamefont {Sheng}, \citenamefont {Wang}, \citenamefont {Luo}, \citenamefont {Wu}, \citenamefont {Wang}, \citenamefont {Hu}, \citenamefont {Xu}, \citenamefont {Sun}, \citenamefont {Zhou}, \citenamefont {Shi}, \citenamefont {Sun}, \citenamefont {Zhang},\ and\ \citenamefont {Bao}}]{Xu2019}%
  \BibitemOpen
  \bibfield  {author} {\bibinfo {author} {\bibfnamefont {H.}~\bibnamefont {Xu}}, \bibinfo {author} {\bibfnamefont {H.}~\bibnamefont {Zhang}}, \bibinfo {author} {\bibfnamefont {Y.}~\bibnamefont {Liu}}, \bibinfo {author} {\bibfnamefont {S.}~\bibnamefont {Zhang}}, \bibinfo {author} {\bibfnamefont {Y.}~\bibnamefont {Sun}}, \bibinfo {author} {\bibfnamefont {Z.}~\bibnamefont {Guo}}, \bibinfo {author} {\bibfnamefont {Y.}~\bibnamefont {Sheng}}, \bibinfo {author} {\bibfnamefont {X.}~\bibnamefont {Wang}}, \bibinfo {author} {\bibfnamefont {C.}~\bibnamefont {Luo}}, \bibinfo {author} {\bibfnamefont {X.}~\bibnamefont {Wu}}, \bibinfo {author} {\bibfnamefont {J.}~\bibnamefont {Wang}}, \bibinfo {author} {\bibfnamefont {W.}~\bibnamefont {Hu}}, \bibinfo {author} {\bibfnamefont {Z.}~\bibnamefont {Xu}}, \bibinfo {author} {\bibfnamefont {Q.}~\bibnamefont {Sun}}, \bibinfo {author} {\bibfnamefont {P.}~\bibnamefont {Zhou}}, \bibinfo {author} {\bibfnamefont {J.}~\bibnamefont {Shi}}, \bibinfo {author} {\bibfnamefont {Z.}~\bibnamefont
  {Sun}}, \bibinfo {author} {\bibfnamefont {D.~W.}\ \bibnamefont {Zhang}}, \ and\ \bibinfo {author} {\bibfnamefont {W.}~\bibnamefont {Bao}},\ }\href {\doibase https://doi.org/10.1002/adfm.201805614} {\bibfield  {journal} {\bibinfo  {journal} {Advanced Functional Materials}\ }\textbf {\bibinfo {volume} {29}},\ \bibinfo {pages} {1805614} (\bibinfo {year} {2019})},\ \Eprint {http://arxiv.org/abs/https://advanced.onlinelibrary.wiley.com/doi/pdf/10.1002/adfm.201805614} {https://advanced.onlinelibrary.wiley.com/doi/pdf/10.1002/adfm.201805614} \BibitemShut {NoStop}%
\bibitem [{\citenamefont {Koperski}\ \emph {et~al.}(2015)\citenamefont {Koperski}, \citenamefont {Nogajewski}, \citenamefont {Arora}, \citenamefont {Cherkez}, \citenamefont {Mallet}, \citenamefont {Veuillen}, \citenamefont {Marcus}, \citenamefont {Kossacki},\ and\ \citenamefont {Potemski}}]{Koperski2015}%
  \BibitemOpen
  \bibfield  {author} {\bibinfo {author} {\bibfnamefont {M.}~\bibnamefont {Koperski}}, \bibinfo {author} {\bibfnamefont {K.}~\bibnamefont {Nogajewski}}, \bibinfo {author} {\bibfnamefont {A.}~\bibnamefont {Arora}}, \bibinfo {author} {\bibfnamefont {V.}~\bibnamefont {Cherkez}}, \bibinfo {author} {\bibfnamefont {P.}~\bibnamefont {Mallet}}, \bibinfo {author} {\bibfnamefont {J.-Y.}\ \bibnamefont {Veuillen}}, \bibinfo {author} {\bibfnamefont {J.}~\bibnamefont {Marcus}}, \bibinfo {author} {\bibfnamefont {P.}~\bibnamefont {Kossacki}}, \ and\ \bibinfo {author} {\bibfnamefont {M.}~\bibnamefont {Potemski}},\ }\href {\doibase 10.1038/nnano.2015.67} {\bibfield  {journal} {\bibinfo  {journal} {Nature Nanotechnology}\ }\textbf {\bibinfo {volume} {10}},\ \bibinfo {pages} {503} (\bibinfo {year} {2015})}\BibitemShut {NoStop}%
\bibitem [{\citenamefont {Withers}\ \emph {et~al.}(2015)\citenamefont {Withers}, \citenamefont {Pozo-Zamudio}, \citenamefont {Mishchenko}, \citenamefont {Rooney}, \citenamefont {Gholinia}, \citenamefont {Watanabe}, \citenamefont {Taniguchi}, \citenamefont {Haigh}, \citenamefont {Geim}, \citenamefont {Tartakovskii},\ and\ \citenamefont {Novoselov}}]{Withers2015}%
  \BibitemOpen
  \bibfield  {author} {\bibinfo {author} {\bibfnamefont {F.}~\bibnamefont {Withers}}, \bibinfo {author} {\bibfnamefont {O.~D.}\ \bibnamefont {Pozo-Zamudio}}, \bibinfo {author} {\bibfnamefont {A.}~\bibnamefont {Mishchenko}}, \bibinfo {author} {\bibfnamefont {A.~P.}\ \bibnamefont {Rooney}}, \bibinfo {author} {\bibfnamefont {A.}~\bibnamefont {Gholinia}}, \bibinfo {author} {\bibfnamefont {K.}~\bibnamefont {Watanabe}}, \bibinfo {author} {\bibfnamefont {T.}~\bibnamefont {Taniguchi}}, \bibinfo {author} {\bibfnamefont {S.~J.}\ \bibnamefont {Haigh}}, \bibinfo {author} {\bibfnamefont {A.~K.}\ \bibnamefont {Geim}}, \bibinfo {author} {\bibfnamefont {A.~I.}\ \bibnamefont {Tartakovskii}}, \ and\ \bibinfo {author} {\bibfnamefont {K.~S.}\ \bibnamefont {Novoselov}},\ }\href {\doibase 10.1038/nmat4205} {\bibfield  {journal} {\bibinfo  {journal} {Nature Materials}\ }\textbf {\bibinfo {volume} {14}},\ \bibinfo {pages} {301} (\bibinfo {year} {2015})}\BibitemShut {NoStop}%
\bibitem [{\citenamefont {Wagner}\ \emph {et~al.}(2018)\citenamefont {Wagner}, \citenamefont {Yim}, \citenamefont {McEvoy}, \citenamefont {Kataria}, \citenamefont {Yokaribas}, \citenamefont {Kuc}, \citenamefont {Pindl}, \citenamefont {Fritzen}, \citenamefont {Heine}, \citenamefont {Duesberg},\ and\ \citenamefont {Lemme}}]{Wagner2018}%
  \BibitemOpen
  \bibfield  {author} {\bibinfo {author} {\bibfnamefont {S.}~\bibnamefont {Wagner}}, \bibinfo {author} {\bibfnamefont {C.}~\bibnamefont {Yim}}, \bibinfo {author} {\bibfnamefont {N.}~\bibnamefont {McEvoy}}, \bibinfo {author} {\bibfnamefont {S.}~\bibnamefont {Kataria}}, \bibinfo {author} {\bibfnamefont {V.}~\bibnamefont {Yokaribas}}, \bibinfo {author} {\bibfnamefont {A.}~\bibnamefont {Kuc}}, \bibinfo {author} {\bibfnamefont {S.}~\bibnamefont {Pindl}}, \bibinfo {author} {\bibfnamefont {C.~P.}\ \bibnamefont {Fritzen}}, \bibinfo {author} {\bibfnamefont {T.}~\bibnamefont {Heine}}, \bibinfo {author} {\bibfnamefont {G.~S.}\ \bibnamefont {Duesberg}}, \ and\ \bibinfo {author} {\bibfnamefont {M.~C.}\ \bibnamefont {Lemme}},\ }\href {\doibase 10.1021/acs.nanolett.8b00928} {\bibfield  {journal} {\bibinfo  {journal} {Nano Letters}\ }\textbf {\bibinfo {volume} {18}},\ \bibinfo {pages} {3738} (\bibinfo {year} {2018})}\BibitemShut {NoStop}%
\bibitem [{\citenamefont {Lopez-Sanchez}\ \emph {et~al.}(2013)\citenamefont {Lopez-Sanchez}, \citenamefont {Lembke}, \citenamefont {Kayci}, \citenamefont {Radenovic},\ and\ \citenamefont {Kis}}]{Lopez-Sanchez2013}%
  \BibitemOpen
  \bibfield  {author} {\bibinfo {author} {\bibfnamefont {O.}~\bibnamefont {Lopez-Sanchez}}, \bibinfo {author} {\bibfnamefont {D.}~\bibnamefont {Lembke}}, \bibinfo {author} {\bibfnamefont {M.}~\bibnamefont {Kayci}}, \bibinfo {author} {\bibfnamefont {A.}~\bibnamefont {Radenovic}}, \ and\ \bibinfo {author} {\bibfnamefont {A.}~\bibnamefont {Kis}},\ }\href {\doibase 10.1038/nnano.2013.100} {\bibfield  {journal} {\bibinfo  {journal} {Nature Nanotechnology}\ }\textbf {\bibinfo {volume} {8}},\ \bibinfo {pages} {497} (\bibinfo {year} {2013})}\BibitemShut {NoStop}%
\bibitem [{\citenamefont {Zeng}\ \emph {et~al.}(2018)\citenamefont {Zeng}, \citenamefont {Lin}, \citenamefont {Lou},\ and\ \citenamefont {et~al.}}]{Zeng2018}%
  \BibitemOpen
  \bibfield  {author} {\bibinfo {author} {\bibfnamefont {L.}~\bibnamefont {Zeng}}, \bibinfo {author} {\bibfnamefont {S.}~\bibnamefont {Lin}}, \bibinfo {author} {\bibfnamefont {Z.}~\bibnamefont {Lou}}, \ and\ \bibinfo {author} {\bibfnamefont {Y.}~\bibnamefont {et~al.}},\ }\href {\doibase 10.1038/s41427-018-0035-4} {\bibfield  {journal} {\bibinfo  {journal} {NPG Asia Materials}\ }\textbf {\bibinfo {volume} {10}},\ \bibinfo {pages} {352} (\bibinfo {year} {2018})}\BibitemShut {NoStop}%
\bibitem [{\citenamefont {Zhu}, \citenamefont {Cheng},\ and\ \citenamefont {Schwingenschl\"ogl}(2011)}]{Zhu_2011}%
  \BibitemOpen
  \bibfield  {author} {\bibinfo {author} {\bibfnamefont {Z.~Y.}\ \bibnamefont {Zhu}}, \bibinfo {author} {\bibfnamefont {Y.~C.}\ \bibnamefont {Cheng}}, \ and\ \bibinfo {author} {\bibfnamefont {U.}~\bibnamefont {Schwingenschl\"ogl}},\ }\href {\doibase 10.1103/PhysRevB.84.153402} {\bibfield  {journal} {\bibinfo  {journal} {Phys. Rev. B}\ }\textbf {\bibinfo {volume} {84}},\ \bibinfo {pages} {153402} (\bibinfo {year} {2011})}\BibitemShut {NoStop}%
\bibitem [{\citenamefont {Mak}\ \emph {et~al.}(2014)\citenamefont {Mak}, \citenamefont {McGill}, \citenamefont {Park},\ and\ \citenamefont {McEuen}}]{Mak2014}%
  \BibitemOpen
  \bibfield  {author} {\bibinfo {author} {\bibfnamefont {K.~F.}\ \bibnamefont {Mak}}, \bibinfo {author} {\bibfnamefont {K.~L.}\ \bibnamefont {McGill}}, \bibinfo {author} {\bibfnamefont {J.}~\bibnamefont {Park}}, \ and\ \bibinfo {author} {\bibfnamefont {P.~L.}\ \bibnamefont {McEuen}},\ }\href {\doibase 10.1126/science.1250140} {\bibfield  {journal} {\bibinfo  {journal} {Science}\ }\textbf {\bibinfo {volume} {344}},\ \bibinfo {pages} {1489} (\bibinfo {year} {2014})}\BibitemShut {NoStop}%
\bibitem [{\citenamefont {Luo}\ \emph {et~al.}(2017)\citenamefont {Luo}, \citenamefont {Xu}, \citenamefont {Zhu}, \citenamefont {Wu}, \citenamefont {McCormick}, \citenamefont {Zhan}, \citenamefont {Neupane},\ and\ \citenamefont {Kawakami}}]{Luo2017}%
  \BibitemOpen
  \bibfield  {author} {\bibinfo {author} {\bibfnamefont {Y.~K.}\ \bibnamefont {Luo}}, \bibinfo {author} {\bibfnamefont {J.}~\bibnamefont {Xu}}, \bibinfo {author} {\bibfnamefont {T.}~\bibnamefont {Zhu}}, \bibinfo {author} {\bibfnamefont {G.}~\bibnamefont {Wu}}, \bibinfo {author} {\bibfnamefont {E.~J.}\ \bibnamefont {McCormick}}, \bibinfo {author} {\bibfnamefont {W.}~\bibnamefont {Zhan}}, \bibinfo {author} {\bibfnamefont {M.~R.}\ \bibnamefont {Neupane}}, \ and\ \bibinfo {author} {\bibfnamefont {R.~K.}\ \bibnamefont {Kawakami}},\ }\href {\doibase 10.1021/acs.nanolett.7b01393} {\bibfield  {journal} {\bibinfo  {journal} {Nano Letters}\ }\textbf {\bibinfo {volume} {17}},\ \bibinfo {pages} {3877} (\bibinfo {year} {2017})}\BibitemShut {NoStop}%
\bibitem [{\citenamefont {Langer}\ \emph {et~al.}(2018)\citenamefont {Langer}, \citenamefont {Schmid}, \citenamefont {Schlauderer}, \citenamefont {Gmitra}, \citenamefont {Fabian}, \citenamefont {Nagler}, \citenamefont {Schüller}, \citenamefont {Korn}, \citenamefont {Hawkins}, \citenamefont {Steiner}, \citenamefont {Huttner}, \citenamefont {Koch}, \citenamefont {Kira},\ and\ \citenamefont {Huber}}]{Langer2018}%
  \BibitemOpen
  \bibfield  {author} {\bibinfo {author} {\bibfnamefont {F.}~\bibnamefont {Langer}}, \bibinfo {author} {\bibfnamefont {C.~P.}\ \bibnamefont {Schmid}}, \bibinfo {author} {\bibfnamefont {S.}~\bibnamefont {Schlauderer}}, \bibinfo {author} {\bibfnamefont {M.}~\bibnamefont {Gmitra}}, \bibinfo {author} {\bibfnamefont {J.}~\bibnamefont {Fabian}}, \bibinfo {author} {\bibfnamefont {P.}~\bibnamefont {Nagler}}, \bibinfo {author} {\bibfnamefont {C.}~\bibnamefont {Schüller}}, \bibinfo {author} {\bibfnamefont {T.}~\bibnamefont {Korn}}, \bibinfo {author} {\bibfnamefont {P.~G.}\ \bibnamefont {Hawkins}}, \bibinfo {author} {\bibfnamefont {J.~T.}\ \bibnamefont {Steiner}}, \bibinfo {author} {\bibfnamefont {U.}~\bibnamefont {Huttner}}, \bibinfo {author} {\bibfnamefont {S.~W.}\ \bibnamefont {Koch}}, \bibinfo {author} {\bibfnamefont {M.}~\bibnamefont {Kira}}, \ and\ \bibinfo {author} {\bibfnamefont {R.}~\bibnamefont {Huber}},\ }\href {\doibase 10.1038/s41586-018-0013-6} {\bibfield  {journal} {\bibinfo  {journal} {Nature}\ }\textbf
  {\bibinfo {volume} {557}},\ \bibinfo {pages} {76} (\bibinfo {year} {2018})}\BibitemShut {NoStop}%
\bibitem [{\citenamefont {Abdukayumov}\ \emph {et~al.}(2024)\citenamefont {Abdukayumov}, \citenamefont {Mičica}, \citenamefont {Ibrahim}, \citenamefont {Vojáček}, \citenamefont {Vergnaud}, \citenamefont {Marty}, \citenamefont {Veuillen}, \citenamefont {Mallet}, \citenamefont {de~Moraes}, \citenamefont {Dosenovic}, \citenamefont {Gambarelli}, \citenamefont {Maurel}, \citenamefont {Wright}, \citenamefont {Tignon}, \citenamefont {Mangeney}, \citenamefont {Ouerghi}, \citenamefont {Renard}, \citenamefont {Mesple}, \citenamefont {Li}, \citenamefont {Bonell}, \citenamefont {Okuno}, \citenamefont {Chshiev}, \citenamefont {George}, \citenamefont {Jaffrès}, \citenamefont {Dhillon},\ and\ \citenamefont {Jamet}}]{Abdukayumov2024}%
  \BibitemOpen
  \bibfield  {author} {\bibinfo {author} {\bibfnamefont {K.}~\bibnamefont {Abdukayumov}}, \bibinfo {author} {\bibfnamefont {M.}~\bibnamefont {Mičica}}, \bibinfo {author} {\bibfnamefont {F.}~\bibnamefont {Ibrahim}}, \bibinfo {author} {\bibfnamefont {L.}~\bibnamefont {Vojáček}}, \bibinfo {author} {\bibfnamefont {C.}~\bibnamefont {Vergnaud}}, \bibinfo {author} {\bibfnamefont {A.}~\bibnamefont {Marty}}, \bibinfo {author} {\bibfnamefont {J.~Y.}\ \bibnamefont {Veuillen}}, \bibinfo {author} {\bibfnamefont {P.}~\bibnamefont {Mallet}}, \bibinfo {author} {\bibfnamefont {I.~G.}\ \bibnamefont {de~Moraes}}, \bibinfo {author} {\bibfnamefont {D.}~\bibnamefont {Dosenovic}}, \bibinfo {author} {\bibfnamefont {S.}~\bibnamefont {Gambarelli}}, \bibinfo {author} {\bibfnamefont {V.}~\bibnamefont {Maurel}}, \bibinfo {author} {\bibfnamefont {A.}~\bibnamefont {Wright}}, \bibinfo {author} {\bibfnamefont {J.}~\bibnamefont {Tignon}}, \bibinfo {author} {\bibfnamefont {J.}~\bibnamefont {Mangeney}}, \bibinfo {author} {\bibfnamefont
  {A.}~\bibnamefont {Ouerghi}}, \bibinfo {author} {\bibfnamefont {V.}~\bibnamefont {Renard}}, \bibinfo {author} {\bibfnamefont {F.}~\bibnamefont {Mesple}}, \bibinfo {author} {\bibfnamefont {J.}~\bibnamefont {Li}}, \bibinfo {author} {\bibfnamefont {F.}~\bibnamefont {Bonell}}, \bibinfo {author} {\bibfnamefont {H.}~\bibnamefont {Okuno}}, \bibinfo {author} {\bibfnamefont {M.}~\bibnamefont {Chshiev}}, \bibinfo {author} {\bibfnamefont {J.~M.}\ \bibnamefont {George}}, \bibinfo {author} {\bibfnamefont {H.}~\bibnamefont {Jaffrès}}, \bibinfo {author} {\bibfnamefont {S.}~\bibnamefont {Dhillon}}, \ and\ \bibinfo {author} {\bibfnamefont {M.}~\bibnamefont {Jamet}},\ }\href {\doibase 10.1002/adma.202304243} {\bibfield  {journal} {\bibinfo  {journal} {Advanced Materials}\ }\textbf {\bibinfo {volume} {36}} (\bibinfo {year} {2024}),\ 10.1002/adma.202304243}\BibitemShut {NoStop}%
\bibitem [{\citenamefont {Mueller}\ and\ \citenamefont {Malic}(2018)}]{Mueller2018}%
  \BibitemOpen
  \bibfield  {author} {\bibinfo {author} {\bibfnamefont {T.}~\bibnamefont {Mueller}}\ and\ \bibinfo {author} {\bibfnamefont {E.}~\bibnamefont {Malic}},\ }\href {\doibase 10.1038/s41699-018-0074-2} {\bibfield  {journal} {\bibinfo  {journal} {npj 2D Materials and Applications}\ }\textbf {\bibinfo {volume} {2}},\ \bibinfo {pages} {1} (\bibinfo {year} {2018})}\BibitemShut {NoStop}%
\bibitem [{\citenamefont {Dutta}\ \emph {et~al.}(2023)\citenamefont {Dutta}, \citenamefont {Bala}, \citenamefont {Sen}, \citenamefont {Spinazze}, \citenamefont {Park}, \citenamefont {Choi}, \citenamefont {Yoon},\ and\ \citenamefont {Kim}}]{Dutta2023}%
  \BibitemOpen
  \bibfield  {author} {\bibinfo {author} {\bibfnamefont {R.}~\bibnamefont {Dutta}}, \bibinfo {author} {\bibfnamefont {A.}~\bibnamefont {Bala}}, \bibinfo {author} {\bibfnamefont {A.}~\bibnamefont {Sen}}, \bibinfo {author} {\bibfnamefont {M.~R.}\ \bibnamefont {Spinazze}}, \bibinfo {author} {\bibfnamefont {H.}~\bibnamefont {Park}}, \bibinfo {author} {\bibfnamefont {W.}~\bibnamefont {Choi}}, \bibinfo {author} {\bibfnamefont {Y.}~\bibnamefont {Yoon}}, \ and\ \bibinfo {author} {\bibfnamefont {S.}~\bibnamefont {Kim}},\ }\href {\doibase 10.1002/adma.202303272} {\bibfield  {journal} {\bibinfo  {journal} {Advanced Materials}\ }\textbf {\bibinfo {volume} {35}},\ \bibinfo {pages} {1} (\bibinfo {year} {2023})}\BibitemShut {NoStop}%
\bibitem [{\citenamefont {Zhang}\ \emph {et~al.}(2014)\citenamefont {Zhang}, \citenamefont {Huang}, \citenamefont {Zhang},\ and\ \citenamefont {Li}}]{Zhang2014}%
  \BibitemOpen
  \bibfield  {author} {\bibinfo {author} {\bibfnamefont {W.}~\bibnamefont {Zhang}}, \bibinfo {author} {\bibfnamefont {Z.}~\bibnamefont {Huang}}, \bibinfo {author} {\bibfnamefont {W.}~\bibnamefont {Zhang}}, \ and\ \bibinfo {author} {\bibfnamefont {Y.}~\bibnamefont {Li}},\ }\href {\doibase 10.1007/s12274-014-0532-x} {\bibfield  {journal} {\bibinfo  {journal} {Nano Research}\ }\textbf {\bibinfo {volume} {7}},\ \bibinfo {pages} {1731} (\bibinfo {year} {2014})}\BibitemShut {NoStop}%
\bibitem [{\citenamefont {Zhao}\ \emph {et~al.}(2017)\citenamefont {Zhao}, \citenamefont {Qiao}, \citenamefont {Yu}, \citenamefont {Yu}, \citenamefont {Xu}, \citenamefont {Lau}, \citenamefont {Zhou}, \citenamefont {Liu}, \citenamefont {Wang}, \citenamefont {Ji},\ and\ \citenamefont {Chai}}]{Zhao2017}%
  \BibitemOpen
  \bibfield  {author} {\bibinfo {author} {\bibfnamefont {Y.}~\bibnamefont {Zhao}}, \bibinfo {author} {\bibfnamefont {J.}~\bibnamefont {Qiao}}, \bibinfo {author} {\bibfnamefont {Z.}~\bibnamefont {Yu}}, \bibinfo {author} {\bibfnamefont {P.}~\bibnamefont {Yu}}, \bibinfo {author} {\bibfnamefont {K.}~\bibnamefont {Xu}}, \bibinfo {author} {\bibfnamefont {S.~P.}\ \bibnamefont {Lau}}, \bibinfo {author} {\bibfnamefont {W.}~\bibnamefont {Zhou}}, \bibinfo {author} {\bibfnamefont {Z.}~\bibnamefont {Liu}}, \bibinfo {author} {\bibfnamefont {X.}~\bibnamefont {Wang}}, \bibinfo {author} {\bibfnamefont {W.}~\bibnamefont {Ji}}, \ and\ \bibinfo {author} {\bibfnamefont {Y.}~\bibnamefont {Chai}},\ }\href {\doibase 10.1002/adma.201604230} {\bibfield  {journal} {\bibinfo  {journal} {Advanced Materials}\ }\textbf {\bibinfo {volume} {29}} (\bibinfo {year} {2017}),\ 10.1002/adma.201604230}\BibitemShut {NoStop}%
\bibitem [{\citenamefont {Yang}\ \emph {et~al.}(2019)\citenamefont {Yang}, \citenamefont {Jang}, \citenamefont {Choi}, \citenamefont {Xu},\ and\ \citenamefont {Lee}}]{Yang2019}%
  \BibitemOpen
  \bibfield  {author} {\bibinfo {author} {\bibfnamefont {Y.}~\bibnamefont {Yang}}, \bibinfo {author} {\bibfnamefont {S.~K.}\ \bibnamefont {Jang}}, \bibinfo {author} {\bibfnamefont {H.}~\bibnamefont {Choi}}, \bibinfo {author} {\bibfnamefont {J.}~\bibnamefont {Xu}}, \ and\ \bibinfo {author} {\bibfnamefont {S.}~\bibnamefont {Lee}},\ }\href {\doibase 10.1039/c9nr07995e} {\bibfield  {journal} {\bibinfo  {journal} {Nanoscale}\ }\textbf {\bibinfo {volume} {11}},\ \bibinfo {pages} {21068} (\bibinfo {year} {2019})}\BibitemShut {NoStop}%
\bibitem [{\citenamefont {Bonell}\ \emph {et~al.}(2021)\citenamefont {Bonell}, \citenamefont {Marty}, \citenamefont {Vergnaud}, \citenamefont {Consonni}, \citenamefont {Okuno}, \citenamefont {Ouerghi}, \citenamefont {Boukari},\ and\ \citenamefont {Jamet}}]{Bonell_2022}%
  \BibitemOpen
  \bibfield  {author} {\bibinfo {author} {\bibfnamefont {F.}~\bibnamefont {Bonell}}, \bibinfo {author} {\bibfnamefont {A.}~\bibnamefont {Marty}}, \bibinfo {author} {\bibfnamefont {C.}~\bibnamefont {Vergnaud}}, \bibinfo {author} {\bibfnamefont {V.}~\bibnamefont {Consonni}}, \bibinfo {author} {\bibfnamefont {H.}~\bibnamefont {Okuno}}, \bibinfo {author} {\bibfnamefont {A.}~\bibnamefont {Ouerghi}}, \bibinfo {author} {\bibfnamefont {H.}~\bibnamefont {Boukari}}, \ and\ \bibinfo {author} {\bibfnamefont {M.}~\bibnamefont {Jamet}},\ }\href {\doibase 10.1088/2053-1583/ac37aa} {\bibfield  {journal} {\bibinfo  {journal} {2D Materials}\ }\textbf {\bibinfo {volume} {9}},\ \bibinfo {pages} {015015} (\bibinfo {year} {2021})}\BibitemShut {NoStop}%
\bibitem [{\citenamefont {Xu}\ \emph {et~al.}(2021{\natexlab{a}})\citenamefont {Xu}, \citenamefont {Wang}, \citenamefont {Liu}, \citenamefont {Li}, \citenamefont {Li}, \citenamefont {Cao}, \citenamefont {Wu}, \citenamefont {Bai},\ and\ \citenamefont {Qi}}]{Xu2021}%
  \BibitemOpen
  \bibfield  {author} {\bibinfo {author} {\bibfnamefont {L.}~\bibnamefont {Xu}}, \bibinfo {author} {\bibfnamefont {L.}~\bibnamefont {Wang}}, \bibinfo {author} {\bibfnamefont {H.}~\bibnamefont {Liu}}, \bibinfo {author} {\bibfnamefont {F.}~\bibnamefont {Li}}, \bibinfo {author} {\bibfnamefont {D.}~\bibnamefont {Li}}, \bibinfo {author} {\bibfnamefont {Y.}~\bibnamefont {Cao}}, \bibinfo {author} {\bibfnamefont {C.}~\bibnamefont {Wu}}, \bibinfo {author} {\bibfnamefont {X.}~\bibnamefont {Bai}}, \ and\ \bibinfo {author} {\bibfnamefont {J.}~\bibnamefont {Qi}},\ }\href {\doibase 10.1039/D0TC05710J} {\bibfield  {journal} {\bibinfo  {journal} {J. Mater. Chem. C}\ }\textbf {\bibinfo {volume} {9}},\ \bibinfo {pages} {5261} (\bibinfo {year} {2021}{\natexlab{a}})}\BibitemShut {NoStop}%
\bibitem [{\citenamefont {Guo}\ and\ \citenamefont {Liang}(1986)}]{Guo_1986}%
  \BibitemOpen
  \bibfield  {author} {\bibinfo {author} {\bibfnamefont {G.~Y.}\ \bibnamefont {Guo}}\ and\ \bibinfo {author} {\bibfnamefont {W.~Y.}\ \bibnamefont {Liang}},\ }\href {\doibase 10.1088/0022-3719/19/7/011} {\bibfield  {journal} {\bibinfo  {journal} {Journal of Physics C: Solid State Physics}\ }\textbf {\bibinfo {volume} {19}},\ \bibinfo {pages} {995} (\bibinfo {year} {1986})}\BibitemShut {NoStop}%
\bibitem [{\citenamefont {Wang}\ \emph {et~al.}(2015)\citenamefont {Wang}, \citenamefont {Li}, \citenamefont {Yao}, \citenamefont {Song}, \citenamefont {Sun}, \citenamefont {Pan}, \citenamefont {Ren}, \citenamefont {Li}, \citenamefont {Okunishi}, \citenamefont {Wang},\ and\ \citenamefont {et~al.}}]{ptse2_2015_wang}%
  \BibitemOpen
  \bibfield  {author} {\bibinfo {author} {\bibfnamefont {Y.}~\bibnamefont {Wang}}, \bibinfo {author} {\bibfnamefont {L.}~\bibnamefont {Li}}, \bibinfo {author} {\bibfnamefont {W.}~\bibnamefont {Yao}}, \bibinfo {author} {\bibfnamefont {S.}~\bibnamefont {Song}}, \bibinfo {author} {\bibfnamefont {J.~T.}\ \bibnamefont {Sun}}, \bibinfo {author} {\bibfnamefont {J.}~\bibnamefont {Pan}}, \bibinfo {author} {\bibfnamefont {X.}~\bibnamefont {Ren}}, \bibinfo {author} {\bibfnamefont {C.}~\bibnamefont {Li}}, \bibinfo {author} {\bibfnamefont {E.}~\bibnamefont {Okunishi}}, \bibinfo {author} {\bibfnamefont {Y.-Q.}\ \bibnamefont {Wang}}, \ and\ \bibinfo {author} {\bibfnamefont {W.}~\bibnamefont {et~al.}},\ }\href {\doibase 10.1021/acs.nanolett.5b00964} {\bibfield  {journal} {\bibinfo  {journal} {Nano Letters}\ }\textbf {\bibinfo {volume} {15}},\ \bibinfo {pages} {4013} (\bibinfo {year} {2015})}\BibitemShut {NoStop}%
\bibitem [{\citenamefont {Yu}\ \emph {et~al.}(2018)\citenamefont {Yu}, \citenamefont {Yu}, \citenamefont {Wu}, \citenamefont {Singh}, \citenamefont {Zeng}, \citenamefont {Lin}, \citenamefont {Zhou}, \citenamefont {Lin}, \citenamefont {Suenaga}, \citenamefont {Liu},\ and\ \citenamefont {Wang}}]{Yu2018}%
  \BibitemOpen
  \bibfield  {author} {\bibinfo {author} {\bibfnamefont {X.}~\bibnamefont {Yu}}, \bibinfo {author} {\bibfnamefont {P.}~\bibnamefont {Yu}}, \bibinfo {author} {\bibfnamefont {D.}~\bibnamefont {Wu}}, \bibinfo {author} {\bibfnamefont {B.}~\bibnamefont {Singh}}, \bibinfo {author} {\bibfnamefont {Q.}~\bibnamefont {Zeng}}, \bibinfo {author} {\bibfnamefont {H.}~\bibnamefont {Lin}}, \bibinfo {author} {\bibfnamefont {W.}~\bibnamefont {Zhou}}, \bibinfo {author} {\bibfnamefont {J.}~\bibnamefont {Lin}}, \bibinfo {author} {\bibfnamefont {K.}~\bibnamefont {Suenaga}}, \bibinfo {author} {\bibfnamefont {Z.}~\bibnamefont {Liu}}, \ and\ \bibinfo {author} {\bibfnamefont {Q.~J.}\ \bibnamefont {Wang}},\ }\href {\doibase 10.1038/s41467-018-03935-0} {\bibfield  {journal} {\bibinfo  {journal} {Nature Communications}\ }\textbf {\bibinfo {volume} {9}},\ \bibinfo {pages} {1} (\bibinfo {year} {2018})}\BibitemShut {NoStop}%
\bibitem [{\citenamefont {Sajjad}, \citenamefont {Singh},\ and\ \citenamefont {Schwingenschlögl}(2018)}]{Sajjad2018}%
  \BibitemOpen
  \bibfield  {author} {\bibinfo {author} {\bibfnamefont {M.}~\bibnamefont {Sajjad}}, \bibinfo {author} {\bibfnamefont {N.}~\bibnamefont {Singh}}, \ and\ \bibinfo {author} {\bibfnamefont {U.}~\bibnamefont {Schwingenschlögl}},\ }\href {\doibase 10.1063/1.5010881} {\bibfield  {journal} {\bibinfo  {journal} {Applied Physics Letters}\ }\textbf {\bibinfo {volume} {112}} (\bibinfo {year} {2018}),\ 10.1063/1.5010881}\BibitemShut {NoStop}%
\bibitem [{\citenamefont {Slušná}\ \emph {et~al.}(2021)\citenamefont {Slušná}, \citenamefont {Vojteková}, \citenamefont {Hrdá}, \citenamefont {Pálková}, \citenamefont {Siffalovic}, \citenamefont {Sojková}, \citenamefont {Végsö}, \citenamefont {Hutár}, \citenamefont {Dobročka}, \citenamefont {Varga},\ and\ \citenamefont {Hulman}}]{pribusova_2021}%
  \BibitemOpen
  \bibfield  {author} {\bibinfo {author} {\bibfnamefont {L.~P.}\ \bibnamefont {Slušná}}, \bibinfo {author} {\bibfnamefont {T.}~\bibnamefont {Vojteková}}, \bibinfo {author} {\bibfnamefont {J.}~\bibnamefont {Hrdá}}, \bibinfo {author} {\bibfnamefont {H.}~\bibnamefont {Pálková}}, \bibinfo {author} {\bibfnamefont {P.}~\bibnamefont {Siffalovic}}, \bibinfo {author} {\bibfnamefont {M.}~\bibnamefont {Sojková}}, \bibinfo {author} {\bibfnamefont {K.}~\bibnamefont {Végsö}}, \bibinfo {author} {\bibfnamefont {P.}~\bibnamefont {Hutár}}, \bibinfo {author} {\bibfnamefont {E.}~\bibnamefont {Dobročka}}, \bibinfo {author} {\bibfnamefont {M.}~\bibnamefont {Varga}}, \ and\ \bibinfo {author} {\bibfnamefont {M.}~\bibnamefont {Hulman}},\ }\href {\doibase 10.1021/acsomega.1c04768} {\bibfield  {journal} {\bibinfo  {journal} {ACS Omega}\ }\textbf {\bibinfo {volume} {6}},\ \bibinfo {pages} {35398} (\bibinfo {year} {2021})}\BibitemShut {NoStop}%
\bibitem [{\citenamefont {Bae}\ \emph {et~al.}(2021)\citenamefont {Bae}, \citenamefont {Nah}, \citenamefont {Lee}, \citenamefont {Sajjad}, \citenamefont {Singh}, \citenamefont {Kang}, \citenamefont {Kim}, \citenamefont {Kim}, \citenamefont {Kim}, \citenamefont {Baik}, \citenamefont {Lee},\ and\ \citenamefont {Sim}}]{Bae2021}%
  \BibitemOpen
  \bibfield  {author} {\bibinfo {author} {\bibfnamefont {S.}~\bibnamefont {Bae}}, \bibinfo {author} {\bibfnamefont {S.}~\bibnamefont {Nah}}, \bibinfo {author} {\bibfnamefont {D.}~\bibnamefont {Lee}}, \bibinfo {author} {\bibfnamefont {M.}~\bibnamefont {Sajjad}}, \bibinfo {author} {\bibfnamefont {N.}~\bibnamefont {Singh}}, \bibinfo {author} {\bibfnamefont {K.~M.}\ \bibnamefont {Kang}}, \bibinfo {author} {\bibfnamefont {S.}~\bibnamefont {Kim}}, \bibinfo {author} {\bibfnamefont {G.~J.}\ \bibnamefont {Kim}}, \bibinfo {author} {\bibfnamefont {J.}~\bibnamefont {Kim}}, \bibinfo {author} {\bibfnamefont {H.}~\bibnamefont {Baik}}, \bibinfo {author} {\bibfnamefont {K.}~\bibnamefont {Lee}}, \ and\ \bibinfo {author} {\bibfnamefont {S.}~\bibnamefont {Sim}},\ }\href {\doibase 10.1002/smll.202103400} {\bibfield  {journal} {\bibinfo  {journal} {Small}\ }\textbf {\bibinfo {volume} {17}},\ \bibinfo {pages} {1} (\bibinfo {year} {2021})}\BibitemShut {NoStop}%
\bibitem [{\citenamefont {Yim}\ \emph {et~al.}(2018)\citenamefont {Yim}, \citenamefont {McEvoy}, \citenamefont {Riazimehr}, \citenamefont {Schneider}, \citenamefont {Gity}, \citenamefont {Monaghan}, \citenamefont {Hurley}, \citenamefont {Lemme},\ and\ \citenamefont {Duesberg}}]{Yim2018}%
  \BibitemOpen
  \bibfield  {author} {\bibinfo {author} {\bibfnamefont {C.}~\bibnamefont {Yim}}, \bibinfo {author} {\bibfnamefont {N.}~\bibnamefont {McEvoy}}, \bibinfo {author} {\bibfnamefont {S.}~\bibnamefont {Riazimehr}}, \bibinfo {author} {\bibfnamefont {D.~S.}\ \bibnamefont {Schneider}}, \bibinfo {author} {\bibfnamefont {F.}~\bibnamefont {Gity}}, \bibinfo {author} {\bibfnamefont {S.}~\bibnamefont {Monaghan}}, \bibinfo {author} {\bibfnamefont {P.~K.}\ \bibnamefont {Hurley}}, \bibinfo {author} {\bibfnamefont {M.~C.}\ \bibnamefont {Lemme}}, \ and\ \bibinfo {author} {\bibfnamefont {G.~S.}\ \bibnamefont {Duesberg}},\ }\href {\doibase 10.1021/acs.nanolett.7b05000} {\bibfield  {journal} {\bibinfo  {journal} {Nano Letters}\ }\textbf {\bibinfo {volume} {18}},\ \bibinfo {pages} {1794} (\bibinfo {year} {2018})}\BibitemShut {NoStop}%
\bibitem [{\citenamefont {Lin}\ \emph {et~al.}(2020)\citenamefont {Lin}, \citenamefont {Xu}, \citenamefont {Cao}, \citenamefont {Zhang}, \citenamefont {Zhou}, \citenamefont {Wang}, \citenamefont {Wan}, \citenamefont {Liu}, \citenamefont {Loh}, \citenamefont {Qiu}, \citenamefont {Bao},\ and\ \citenamefont {Jia}}]{Lin2020}%
  \BibitemOpen
  \bibfield  {author} {\bibinfo {author} {\bibfnamefont {H.}~\bibnamefont {Lin}}, \bibinfo {author} {\bibfnamefont {Z.-Q.}\ \bibnamefont {Xu}}, \bibinfo {author} {\bibfnamefont {G.}~\bibnamefont {Cao}}, \bibinfo {author} {\bibfnamefont {Y.}~\bibnamefont {Zhang}}, \bibinfo {author} {\bibfnamefont {J.}~\bibnamefont {Zhou}}, \bibinfo {author} {\bibfnamefont {Z.}~\bibnamefont {Wang}}, \bibinfo {author} {\bibfnamefont {Z.}~\bibnamefont {Wan}}, \bibinfo {author} {\bibfnamefont {Z.}~\bibnamefont {Liu}}, \bibinfo {author} {\bibfnamefont {K.~P.}\ \bibnamefont {Loh}}, \bibinfo {author} {\bibfnamefont {C.-W.}\ \bibnamefont {Qiu}}, \bibinfo {author} {\bibfnamefont {Q.}~\bibnamefont {Bao}}, \ and\ \bibinfo {author} {\bibfnamefont {B.}~\bibnamefont {Jia}},\ }\href {\doibase 10.1038/s41377-020-00374-9} {\bibfield  {journal} {\bibinfo  {journal} {Light: Science \& Applications}\ }\textbf {\bibinfo {volume} {9}},\ \bibinfo {pages} {137} (\bibinfo {year} {2020})}\BibitemShut {NoStop}%
\bibitem [{\citenamefont {Li}, \citenamefont {Li},\ and\ \citenamefont {Zeng}(2016{\natexlab{a}})}]{Li_JMCC_2016}%
  \BibitemOpen
  \bibfield  {author} {\bibinfo {author} {\bibfnamefont {P.}~\bibnamefont {Li}}, \bibinfo {author} {\bibfnamefont {L.}~\bibnamefont {Li}}, \ and\ \bibinfo {author} {\bibfnamefont {X.~C.}\ \bibnamefont {Zeng}},\ }\href {\doibase 10.1039/C6TC00130K} {\bibfield  {journal} {\bibinfo  {journal} {J. Mater. Chem. C}\ }\textbf {\bibinfo {volume} {4}},\ \bibinfo {pages} {3106} (\bibinfo {year} {2016}{\natexlab{a}})}\BibitemShut {NoStop}%
\bibitem [{\citenamefont {Li}, \citenamefont {Li},\ and\ \citenamefont {Zeng}(2016{\natexlab{b}})}]{Li2016}%
  \BibitemOpen
  \bibfield  {author} {\bibinfo {author} {\bibfnamefont {P.}~\bibnamefont {Li}}, \bibinfo {author} {\bibfnamefont {L.}~\bibnamefont {Li}}, \ and\ \bibinfo {author} {\bibfnamefont {X.~C.}\ \bibnamefont {Zeng}},\ }\href {\doibase 10.1039/c6tc00130k} {\bibfield  {journal} {\bibinfo  {journal} {Journal of Materials Chemistry C}\ }\textbf {\bibinfo {volume} {4}},\ \bibinfo {pages} {3106} (\bibinfo {year} {2016}{\natexlab{b}})}\BibitemShut {NoStop}%
\bibitem [{\citenamefont {Deng}, \citenamefont {Li},\ and\ \citenamefont {Zhang}(2018)}]{Deng2018}%
  \BibitemOpen
  \bibfield  {author} {\bibinfo {author} {\bibfnamefont {S.}~\bibnamefont {Deng}}, \bibinfo {author} {\bibfnamefont {L.}~\bibnamefont {Li}}, \ and\ \bibinfo {author} {\bibfnamefont {Y.}~\bibnamefont {Zhang}},\ }\href {\doibase 10.1021/acsanm.8b00363} {\bibfield  {journal} {\bibinfo  {journal} {ACS Applied Nano Materials}\ }\textbf {\bibinfo {volume} {1}},\ \bibinfo {pages} {1932} (\bibinfo {year} {2018})}\BibitemShut {NoStop}%
\bibitem [{\citenamefont {Amairi}, \citenamefont {Smiri},\ and\ \citenamefont {Jaziri}(2024)}]{Amairi_2025}%
  \BibitemOpen
  \bibfield  {author} {\bibinfo {author} {\bibfnamefont {R.}~\bibnamefont {Amairi}}, \bibinfo {author} {\bibfnamefont {A.}~\bibnamefont {Smiri}}, \ and\ \bibinfo {author} {\bibfnamefont {S.}~\bibnamefont {Jaziri}},\ }\href {\doibase 10.1088/1361-648X/ad8697} {\bibfield  {journal} {\bibinfo  {journal} {Journal of Physics: Condensed Matter}\ }\textbf {\bibinfo {volume} {37}},\ \bibinfo {pages} {035501} (\bibinfo {year} {2024})}\BibitemShut {NoStop}%
\bibitem [{\citenamefont {Ciarrocchi}\ \emph {et~al.}(2018)\citenamefont {Ciarrocchi}, \citenamefont {Avsar}, \citenamefont {Ovchinnikov},\ and\ \citenamefont {Kis}}]{Ciarrocchi2018}%
  \BibitemOpen
  \bibfield  {author} {\bibinfo {author} {\bibfnamefont {A.}~\bibnamefont {Ciarrocchi}}, \bibinfo {author} {\bibfnamefont {A.}~\bibnamefont {Avsar}}, \bibinfo {author} {\bibfnamefont {D.}~\bibnamefont {Ovchinnikov}}, \ and\ \bibinfo {author} {\bibfnamefont {A.}~\bibnamefont {Kis}},\ }\href {\doibase 10.1038/s41467-018-03436-0} {\bibfield  {journal} {\bibinfo  {journal} {Nature Communications}\ }\textbf {\bibinfo {volume} {9}},\ \bibinfo {pages} {1} (\bibinfo {year} {2018})}\BibitemShut {NoStop}%
\bibitem [{\citenamefont {Villaos}\ \emph {et~al.}(2019)\citenamefont {Villaos}, \citenamefont {Crisostomo}, \citenamefont {Huang}, \citenamefont {Huang}, \citenamefont {Padama}, \citenamefont {Albao}, \citenamefont {Lin},\ and\ \citenamefont {Chuang}}]{multilayers_PtSe2_Villaos2019}%
  \BibitemOpen
  \bibfield  {author} {\bibinfo {author} {\bibfnamefont {R.~A.~B.}\ \bibnamefont {Villaos}}, \bibinfo {author} {\bibfnamefont {C.~P.}\ \bibnamefont {Crisostomo}}, \bibinfo {author} {\bibfnamefont {Z.~Q.}\ \bibnamefont {Huang}}, \bibinfo {author} {\bibfnamefont {S.~M.}\ \bibnamefont {Huang}}, \bibinfo {author} {\bibfnamefont {A.~A.~B.}\ \bibnamefont {Padama}}, \bibinfo {author} {\bibfnamefont {M.~A.}\ \bibnamefont {Albao}}, \bibinfo {author} {\bibfnamefont {H.}~\bibnamefont {Lin}}, \ and\ \bibinfo {author} {\bibfnamefont {F.~C.}\ \bibnamefont {Chuang}},\ }\href {\doibase 10.1038/s41699-018-0085-z} {\bibfield  {journal} {\bibinfo  {journal} {npj 2D Materials and Applications}\ }\textbf {\bibinfo {volume} {3}},\ \bibinfo {pages} {1} (\bibinfo {year} {2019})}\BibitemShut {NoStop}%
\bibitem [{\citenamefont {Li}\ \emph {et~al.}(2021)\citenamefont {Li}, \citenamefont {Kolekar}, \citenamefont {Ghorbani-Asl}, \citenamefont {Lehnert}, \citenamefont {Biskupek}, \citenamefont {Kaiser}, \citenamefont {Krasheninnikov},\ and\ \citenamefont {Batzill}}]{structural_parameters_Li2021}%
  \BibitemOpen
  \bibfield  {author} {\bibinfo {author} {\bibfnamefont {J.}~\bibnamefont {Li}}, \bibinfo {author} {\bibfnamefont {S.}~\bibnamefont {Kolekar}}, \bibinfo {author} {\bibfnamefont {M.}~\bibnamefont {Ghorbani-Asl}}, \bibinfo {author} {\bibfnamefont {T.}~\bibnamefont {Lehnert}}, \bibinfo {author} {\bibfnamefont {J.}~\bibnamefont {Biskupek}}, \bibinfo {author} {\bibfnamefont {U.}~\bibnamefont {Kaiser}}, \bibinfo {author} {\bibfnamefont {A.~V.}\ \bibnamefont {Krasheninnikov}}, \ and\ \bibinfo {author} {\bibfnamefont {M.}~\bibnamefont {Batzill}},\ }\href {\doibase 10.1021/acsnano.1c02971} {\bibfield  {journal} {\bibinfo  {journal} {ACS Nano}\ }\textbf {\bibinfo {volume} {15}},\ \bibinfo {pages} {13249} (\bibinfo {year} {2021})}\BibitemShut {NoStop}%
\bibitem [{\citenamefont {Zheng}\ \emph {et~al.}(2022)\citenamefont {Zheng}, \citenamefont {Peng}, \citenamefont {Yu}, \citenamefont {Lan}, \citenamefont {Wang}, \citenamefont {Zhang}, \citenamefont {Li}, \citenamefont {Liang},\ and\ \citenamefont {Su}}]{Zheng2022}%
  \BibitemOpen
  \bibfield  {author} {\bibinfo {author} {\bibfnamefont {Z.~S.}\ \bibnamefont {Zheng}}, \bibinfo {author} {\bibfnamefont {Z.}~\bibnamefont {Peng}}, \bibinfo {author} {\bibfnamefont {Z.~S.}\ \bibnamefont {Yu}}, \bibinfo {author} {\bibfnamefont {H.}~\bibnamefont {Lan}}, \bibinfo {author} {\bibfnamefont {S.~X.}\ \bibnamefont {Wang}}, \bibinfo {author} {\bibfnamefont {M.}~\bibnamefont {Zhang}}, \bibinfo {author} {\bibfnamefont {L.}~\bibnamefont {Li}}, \bibinfo {author} {\bibfnamefont {H.~W.}\ \bibnamefont {Liang}}, \ and\ \bibinfo {author} {\bibfnamefont {H.}~\bibnamefont {Su}},\ }\href {\doibase 10.1016/j.rinp.2022.106012} {\bibfield  {journal} {\bibinfo  {journal} {Results in Physics}\ }\textbf {\bibinfo {volume} {42}},\ \bibinfo {pages} {106012} (\bibinfo {year} {2022})}\BibitemShut {NoStop}%
\bibitem [{\citenamefont {Fang}\ \emph {et~al.}(2019)\citenamefont {Fang}, \citenamefont {Liang}, \citenamefont {Feng},\ and\ \citenamefont {Luo}}]{Fang2019}%
  \BibitemOpen
  \bibfield  {author} {\bibinfo {author} {\bibfnamefont {L.}~\bibnamefont {Fang}}, \bibinfo {author} {\bibfnamefont {W.}~\bibnamefont {Liang}}, \bibinfo {author} {\bibfnamefont {Q.}~\bibnamefont {Feng}}, \ and\ \bibinfo {author} {\bibfnamefont {S.~N.}\ \bibnamefont {Luo}},\ }\href {\doibase 10.1088/1361-648X/ab34bc} {\bibfield  {journal} {\bibinfo  {journal} {Journal of Physics Condensed Matter}\ }\textbf {\bibinfo {volume} {31}} (\bibinfo {year} {2019}),\ 10.1088/1361-648X/ab34bc}\BibitemShut {NoStop}%
\bibitem [{\citenamefont {Carr}\ \emph {et~al.}(2017)\citenamefont {Carr}, \citenamefont {Massatt}, \citenamefont {Fang}, \citenamefont {Cazeaux}, \citenamefont {Luskin},\ and\ \citenamefont {Kaxiras}}]{Carr2017}%
  \BibitemOpen
  \bibfield  {author} {\bibinfo {author} {\bibfnamefont {S.}~\bibnamefont {Carr}}, \bibinfo {author} {\bibfnamefont {D.}~\bibnamefont {Massatt}}, \bibinfo {author} {\bibfnamefont {S.}~\bibnamefont {Fang}}, \bibinfo {author} {\bibfnamefont {P.}~\bibnamefont {Cazeaux}}, \bibinfo {author} {\bibfnamefont {M.}~\bibnamefont {Luskin}}, \ and\ \bibinfo {author} {\bibfnamefont {E.}~\bibnamefont {Kaxiras}},\ }\href {\doibase 10.1103/PhysRevB.95.075420} {\bibfield  {journal} {\bibinfo  {journal} {Physical Review B}\ }\textbf {\bibinfo {volume} {95}},\ \bibinfo {pages} {1} (\bibinfo {year} {2017})}\BibitemShut {NoStop}%
\bibitem [{\citenamefont {Li}\ \emph {et~al.}(2010)\citenamefont {Li}, \citenamefont {Luican}, \citenamefont {Lopes~dos Santos}, \citenamefont {Castro~Neto}, \citenamefont {Reina}, \citenamefont {Kong},\ and\ \citenamefont {Andrei}}]{Li2010}%
  \BibitemOpen
  \bibfield  {author} {\bibinfo {author} {\bibfnamefont {G.}~\bibnamefont {Li}}, \bibinfo {author} {\bibfnamefont {A.}~\bibnamefont {Luican}}, \bibinfo {author} {\bibfnamefont {J.~M.~B.}\ \bibnamefont {Lopes~dos Santos}}, \bibinfo {author} {\bibfnamefont {A.~H.}\ \bibnamefont {Castro~Neto}}, \bibinfo {author} {\bibfnamefont {A.}~\bibnamefont {Reina}}, \bibinfo {author} {\bibfnamefont {J.}~\bibnamefont {Kong}}, \ and\ \bibinfo {author} {\bibfnamefont {E.~Y.}\ \bibnamefont {Andrei}},\ }\href {\doibase 10.1038/nphys1463} {\bibfield  {journal} {\bibinfo  {journal} {Nature Physics}\ }\textbf {\bibinfo {volume} {6}},\ \bibinfo {pages} {109} (\bibinfo {year} {2010})}\BibitemShut {NoStop}%
\bibitem [{\citenamefont {Su\'arez~Morell}\ \emph {et~al.}(2010)\citenamefont {Su\'arez~Morell}, \citenamefont {Correa}, \citenamefont {Vargas}, \citenamefont {Pacheco},\ and\ \citenamefont {Barticevic}}]{Morell2010}%
  \BibitemOpen
  \bibfield  {author} {\bibinfo {author} {\bibfnamefont {E.}~\bibnamefont {Su\'arez~Morell}}, \bibinfo {author} {\bibfnamefont {J.~D.}\ \bibnamefont {Correa}}, \bibinfo {author} {\bibfnamefont {P.}~\bibnamefont {Vargas}}, \bibinfo {author} {\bibfnamefont {M.}~\bibnamefont {Pacheco}}, \ and\ \bibinfo {author} {\bibfnamefont {Z.}~\bibnamefont {Barticevic}},\ }\href {\doibase 10.1103/PhysRevB.82.121407} {\bibfield  {journal} {\bibinfo  {journal} {Phys. Rev. B}\ }\textbf {\bibinfo {volume} {82}},\ \bibinfo {pages} {121407} (\bibinfo {year} {2010})}\BibitemShut {NoStop}%
\bibitem [{\citenamefont {Dean}\ \emph {et~al.}(2013)\citenamefont {Dean}, \citenamefont {Wang}, \citenamefont {Maher}, \citenamefont {Forsythe}, \citenamefont {Ghahari}, \citenamefont {Gao}, \citenamefont {Katoch}, \citenamefont {Ishigami}, \citenamefont {Moon}, \citenamefont {Koshino}, \citenamefont {Taniguchi}, \citenamefont {Watanabe}, \citenamefont {Shepard}, \citenamefont {Hone},\ and\ \citenamefont {Kim}}]{Dean2013}%
  \BibitemOpen
  \bibfield  {author} {\bibinfo {author} {\bibfnamefont {C.~R.}\ \bibnamefont {Dean}}, \bibinfo {author} {\bibfnamefont {L.}~\bibnamefont {Wang}}, \bibinfo {author} {\bibfnamefont {P.}~\bibnamefont {Maher}}, \bibinfo {author} {\bibfnamefont {C.}~\bibnamefont {Forsythe}}, \bibinfo {author} {\bibfnamefont {F.}~\bibnamefont {Ghahari}}, \bibinfo {author} {\bibfnamefont {Y.}~\bibnamefont {Gao}}, \bibinfo {author} {\bibfnamefont {J.}~\bibnamefont {Katoch}}, \bibinfo {author} {\bibfnamefont {M.}~\bibnamefont {Ishigami}}, \bibinfo {author} {\bibfnamefont {P.}~\bibnamefont {Moon}}, \bibinfo {author} {\bibfnamefont {M.}~\bibnamefont {Koshino}}, \bibinfo {author} {\bibfnamefont {T.}~\bibnamefont {Taniguchi}}, \bibinfo {author} {\bibfnamefont {K.}~\bibnamefont {Watanabe}}, \bibinfo {author} {\bibfnamefont {K.~L.}\ \bibnamefont {Shepard}}, \bibinfo {author} {\bibfnamefont {J.}~\bibnamefont {Hone}}, \ and\ \bibinfo {author} {\bibfnamefont {P.}~\bibnamefont {Kim}},\ }\href {\doibase 10.1038/nature12186} {\bibfield
  {journal} {\bibinfo  {journal} {Nature}\ }\textbf {\bibinfo {volume} {497}},\ \bibinfo {pages} {598} (\bibinfo {year} {2013})}\BibitemShut {NoStop}%
\bibitem [{\citenamefont {Cao}\ \emph {et~al.}(2018)\citenamefont {Cao}, \citenamefont {Fatemi}, \citenamefont {Fang}, \citenamefont {Watanabe}, \citenamefont {Taniguchi}, \citenamefont {Kaxiras},\ and\ \citenamefont {Jarillo-herrero}}]{Cao2018}%
  \BibitemOpen
  \bibfield  {author} {\bibinfo {author} {\bibfnamefont {Y.}~\bibnamefont {Cao}}, \bibinfo {author} {\bibfnamefont {V.}~\bibnamefont {Fatemi}}, \bibinfo {author} {\bibfnamefont {S.}~\bibnamefont {Fang}}, \bibinfo {author} {\bibfnamefont {K.}~\bibnamefont {Watanabe}}, \bibinfo {author} {\bibfnamefont {T.}~\bibnamefont {Taniguchi}}, \bibinfo {author} {\bibfnamefont {E.}~\bibnamefont {Kaxiras}}, \ and\ \bibinfo {author} {\bibfnamefont {P.}~\bibnamefont {Jarillo-herrero}},\ }\href {\doibase 10.1038/nature26160} {\bibfield  {journal} {\bibinfo  {journal} {Nature Publishing Group}\ } (\bibinfo {year} {2018}),\ 10.1038/nature26160}\BibitemShut {NoStop}%
\bibitem [{\citenamefont {Klebl}\ \emph {et~al.}(2022)\citenamefont {Klebl}, \citenamefont {Xu}, \citenamefont {Fischer}, \citenamefont {Xian}, \citenamefont {Claassen}, \citenamefont {Rubio},\ and\ \citenamefont {Kennes}}]{Klebl2022}%
  \BibitemOpen
  \bibfield  {author} {\bibinfo {author} {\bibfnamefont {L.}~\bibnamefont {Klebl}}, \bibinfo {author} {\bibfnamefont {Q.}~\bibnamefont {Xu}}, \bibinfo {author} {\bibfnamefont {A.}~\bibnamefont {Fischer}}, \bibinfo {author} {\bibfnamefont {L.}~\bibnamefont {Xian}}, \bibinfo {author} {\bibfnamefont {M.}~\bibnamefont {Claassen}}, \bibinfo {author} {\bibfnamefont {A.}~\bibnamefont {Rubio}}, \ and\ \bibinfo {author} {\bibfnamefont {D.~M.}\ \bibnamefont {Kennes}},\ }\href {\doibase 10.1088/2516-1075/ac49f5} {\bibfield  {journal} {\bibinfo  {journal} {Electronic Structure}\ }\textbf {\bibinfo {volume} {4}} (\bibinfo {year} {2022}),\ 10.1088/2516-1075/ac49f5}\BibitemShut {NoStop}%
\bibitem [{\citenamefont {Rickhaus}\ \emph {et~al.}(2020)\citenamefont {Rickhaus}, \citenamefont {Liu}, \citenamefont {Kurpas}, \citenamefont {Kurzmann}, \citenamefont {Lee}, \citenamefont {Overweg}, \citenamefont {Eich}, \citenamefont {Pisoni}, \citenamefont {Taniguchi}, \citenamefont {Watanabe}, \citenamefont {Richter}, \citenamefont {Ensslin},\ and\ \citenamefont {Ihn}}]{Rickhaus2020}%
  \BibitemOpen
  \bibfield  {author} {\bibinfo {author} {\bibfnamefont {P.}~\bibnamefont {Rickhaus}}, \bibinfo {author} {\bibfnamefont {M.-H.}\ \bibnamefont {Liu}}, \bibinfo {author} {\bibfnamefont {M.}~\bibnamefont {Kurpas}}, \bibinfo {author} {\bibfnamefont {A.}~\bibnamefont {Kurzmann}}, \bibinfo {author} {\bibfnamefont {Y.}~\bibnamefont {Lee}}, \bibinfo {author} {\bibfnamefont {H.}~\bibnamefont {Overweg}}, \bibinfo {author} {\bibfnamefont {M.}~\bibnamefont {Eich}}, \bibinfo {author} {\bibfnamefont {R.}~\bibnamefont {Pisoni}}, \bibinfo {author} {\bibfnamefont {T.}~\bibnamefont {Taniguchi}}, \bibinfo {author} {\bibfnamefont {K.}~\bibnamefont {Watanabe}}, \bibinfo {author} {\bibfnamefont {K.}~\bibnamefont {Richter}}, \bibinfo {author} {\bibfnamefont {K.}~\bibnamefont {Ensslin}}, \ and\ \bibinfo {author} {\bibfnamefont {T.}~\bibnamefont {Ihn}},\ }\href {\doibase 10.1126/sciadv.aay8409} {\bibfield  {journal} {\bibinfo  {journal} {Science Advances}\ }\textbf {\bibinfo {volume} {6}},\ \bibinfo {pages} {eaay8409} (\bibinfo
  {year} {2020})},\ \Eprint {http://arxiv.org/abs/https://www.science.org/doi/pdf/10.1126/sciadv.aay8409} {https://www.science.org/doi/pdf/10.1126/sciadv.aay8409} \BibitemShut {NoStop}%
\bibitem [{\citenamefont {Wu}, \citenamefont {Lovorn},\ and\ \citenamefont {MacDonald}(2017)}]{Wu2017}%
  \BibitemOpen
  \bibfield  {author} {\bibinfo {author} {\bibfnamefont {F.}~\bibnamefont {Wu}}, \bibinfo {author} {\bibfnamefont {T.}~\bibnamefont {Lovorn}}, \ and\ \bibinfo {author} {\bibfnamefont {A.~H.}\ \bibnamefont {MacDonald}},\ }\href {\doibase 10.1103/PhysRevLett.118.147401} {\bibfield  {journal} {\bibinfo  {journal} {Phys. Rev. Lett.}\ }\textbf {\bibinfo {volume} {118}},\ \bibinfo {pages} {147401} (\bibinfo {year} {2017})}\BibitemShut {NoStop}%
\bibitem [{\citenamefont {Tran}\ \emph {et~al.}(2019)\citenamefont {Tran}, \citenamefont {Moody}, \citenamefont {Wu}, \citenamefont {Lu}, \citenamefont {Choi}, \citenamefont {Kim}, \citenamefont {Rai}, \citenamefont {Sanchez}, \citenamefont {Quan}, \citenamefont {Singh}, \citenamefont {Embley}, \citenamefont {Zepeda}, \citenamefont {Campbell}, \citenamefont {Autry}, \citenamefont {Taniguchi}, \citenamefont {Watanabe}, \citenamefont {Lu}, \citenamefont {Banerjee}, \citenamefont {Silverman}, \citenamefont {Kim}, \citenamefont {Tutuc}, \citenamefont {Yang}, \citenamefont {MacDonald},\ and\ \citenamefont {Li}}]{Tran2019}%
  \BibitemOpen
  \bibfield  {author} {\bibinfo {author} {\bibfnamefont {K.}~\bibnamefont {Tran}}, \bibinfo {author} {\bibfnamefont {G.}~\bibnamefont {Moody}}, \bibinfo {author} {\bibfnamefont {F.}~\bibnamefont {Wu}}, \bibinfo {author} {\bibfnamefont {X.}~\bibnamefont {Lu}}, \bibinfo {author} {\bibfnamefont {J.}~\bibnamefont {Choi}}, \bibinfo {author} {\bibfnamefont {K.}~\bibnamefont {Kim}}, \bibinfo {author} {\bibfnamefont {A.}~\bibnamefont {Rai}}, \bibinfo {author} {\bibfnamefont {D.~A.}\ \bibnamefont {Sanchez}}, \bibinfo {author} {\bibfnamefont {J.}~\bibnamefont {Quan}}, \bibinfo {author} {\bibfnamefont {A.}~\bibnamefont {Singh}}, \bibinfo {author} {\bibfnamefont {J.}~\bibnamefont {Embley}}, \bibinfo {author} {\bibfnamefont {A.}~\bibnamefont {Zepeda}}, \bibinfo {author} {\bibfnamefont {M.}~\bibnamefont {Campbell}}, \bibinfo {author} {\bibfnamefont {T.}~\bibnamefont {Autry}}, \bibinfo {author} {\bibfnamefont {T.}~\bibnamefont {Taniguchi}}, \bibinfo {author} {\bibfnamefont {K.}~\bibnamefont {Watanabe}}, \bibinfo {author}
  {\bibfnamefont {N.}~\bibnamefont {Lu}}, \bibinfo {author} {\bibfnamefont {S.~K.}\ \bibnamefont {Banerjee}}, \bibinfo {author} {\bibfnamefont {K.~L.}\ \bibnamefont {Silverman}}, \bibinfo {author} {\bibfnamefont {S.}~\bibnamefont {Kim}}, \bibinfo {author} {\bibfnamefont {E.}~\bibnamefont {Tutuc}}, \bibinfo {author} {\bibfnamefont {L.}~\bibnamefont {Yang}}, \bibinfo {author} {\bibfnamefont {A.~H.}\ \bibnamefont {MacDonald}}, \ and\ \bibinfo {author} {\bibfnamefont {X.}~\bibnamefont {Li}},\ }\href {\doibase 10.1038/s41586-019-0975-z} {\bibfield  {journal} {\bibinfo  {journal} {Nature}\ }\textbf {\bibinfo {volume} {567}},\ \bibinfo {pages} {71} (\bibinfo {year} {2019})}\BibitemShut {NoStop}%
\bibitem [{\citenamefont {He}\ \emph {et~al.}(2021)\citenamefont {He}, \citenamefont {Li}, \citenamefont {Li}, \citenamefont {Mao}, \citenamefont {Liu}, \citenamefont {Teng}, \citenamefont {Wang},\ and\ \citenamefont {Wang}}]{He2021}%
  \BibitemOpen
  \bibfield  {author} {\bibinfo {author} {\bibfnamefont {F.}~\bibnamefont {He}}, \bibinfo {author} {\bibfnamefont {J.}~\bibnamefont {Li}}, \bibinfo {author} {\bibfnamefont {L.}~\bibnamefont {Li}}, \bibinfo {author} {\bibfnamefont {X.}~\bibnamefont {Mao}}, \bibinfo {author} {\bibfnamefont {Z.}~\bibnamefont {Liu}}, \bibinfo {author} {\bibfnamefont {S.}~\bibnamefont {Teng}}, \bibinfo {author} {\bibfnamefont {J.}~\bibnamefont {Wang}}, \ and\ \bibinfo {author} {\bibfnamefont {Y.}~\bibnamefont {Wang}},\ }\href {\doibase 10.1209/0295-5075/ac35b9} {\bibfield  {journal} {\bibinfo  {journal} {Epl}\ }\textbf {\bibinfo {volume} {136}} (\bibinfo {year} {2021}),\ 10.1209/0295-5075/ac35b9}\BibitemShut {NoStop}%
\bibitem [{\citenamefont {Huang}\ \emph {et~al.}(2022)\citenamefont {Huang}, \citenamefont {Choi}, \citenamefont {Shih},\ and\ \citenamefont {Li}}]{Huang2022}%
  \BibitemOpen
  \bibfield  {author} {\bibinfo {author} {\bibfnamefont {D.}~\bibnamefont {Huang}}, \bibinfo {author} {\bibfnamefont {J.}~\bibnamefont {Choi}}, \bibinfo {author} {\bibfnamefont {C.-K.}\ \bibnamefont {Shih}}, \ and\ \bibinfo {author} {\bibfnamefont {X.}~\bibnamefont {Li}},\ }\href {\doibase 10.1038/s41565-021-01068-y} {\bibfield  {journal} {\bibinfo  {journal} {Nature Nanotechnology}\ }\textbf {\bibinfo {volume} {17}},\ \bibinfo {pages} {227} (\bibinfo {year} {2022})}\BibitemShut {NoStop}%
\bibitem [{\citenamefont {Gao}\ \emph {et~al.}(2020)\citenamefont {Gao}, \citenamefont {Li}, \citenamefont {Xin}, \citenamefont {Chen}, \citenamefont {Liu},\ and\ \citenamefont {Tian}}]{Gao2020}%
  \BibitemOpen
  \bibfield  {author} {\bibinfo {author} {\bibfnamefont {X.~G.}\ \bibnamefont {Gao}}, \bibinfo {author} {\bibfnamefont {X.~K.}\ \bibnamefont {Li}}, \bibinfo {author} {\bibfnamefont {W.}~\bibnamefont {Xin}}, \bibinfo {author} {\bibfnamefont {X.~D.}\ \bibnamefont {Chen}}, \bibinfo {author} {\bibfnamefont {Z.~B.}\ \bibnamefont {Liu}}, \ and\ \bibinfo {author} {\bibfnamefont {J.~G.}\ \bibnamefont {Tian}},\ }\href {\doibase 10.1515/nanoph-2020-0024} {\bibfield  {journal} {\bibinfo  {journal} {Nanophotonics}\ }\textbf {\bibinfo {volume} {9}},\ \bibinfo {pages} {1717} (\bibinfo {year} {2020})}\BibitemShut {NoStop}%
\bibitem [{\citenamefont {Xin}\ \emph {et~al.}(2022)\citenamefont {Xin}, \citenamefont {Wang}, \citenamefont {Grove-Rasmussen},\ and\ \citenamefont {Wei}}]{Xin2022}%
  \BibitemOpen
  \bibfield  {author} {\bibinfo {author} {\bibfnamefont {K.}~\bibnamefont {Xin}}, \bibinfo {author} {\bibfnamefont {X.}~\bibnamefont {Wang}}, \bibinfo {author} {\bibfnamefont {K.}~\bibnamefont {Grove-Rasmussen}}, \ and\ \bibinfo {author} {\bibfnamefont {Z.}~\bibnamefont {Wei}},\ }\href {\doibase 10.1088/1674-4926/43/1/011001} {\bibfield  {journal} {\bibinfo  {journal} {Journal of Semiconductors}\ }\textbf {\bibinfo {volume} {43}} (\bibinfo {year} {2022}),\ 10.1088/1674-4926/43/1/011001}\BibitemShut {NoStop}%
\bibitem [{\citenamefont {Gupta}\ \emph {et~al.}(2024)\citenamefont {Gupta}, \citenamefont {Sachin}, \citenamefont {Kumari}, \citenamefont {Rani},\ and\ \citenamefont {Ray}}]{Gupta2024}%
  \BibitemOpen
  \bibfield  {author} {\bibinfo {author} {\bibfnamefont {N.}~\bibnamefont {Gupta}}, \bibinfo {author} {\bibfnamefont {S.}~\bibnamefont {Sachin}}, \bibinfo {author} {\bibfnamefont {P.}~\bibnamefont {Kumari}}, \bibinfo {author} {\bibfnamefont {S.}~\bibnamefont {Rani}}, \ and\ \bibinfo {author} {\bibfnamefont {S.~J.}\ \bibnamefont {Ray}},\ }\href {\doibase 10.1039/D3RA06559F} {\bibfield  {journal} {\bibinfo  {journal} {RSC Advances}\ }\textbf {\bibinfo {volume} {14}},\ \bibinfo {pages} {2878} (\bibinfo {year} {2024})}\BibitemShut {NoStop}%
\bibitem [{\citenamefont {Xu}\ \emph {et~al.}(2021{\natexlab{b}})\citenamefont {Xu}, \citenamefont {Liu}, \citenamefont {Song}, \citenamefont {Li}, \citenamefont {Li}, \citenamefont {Li}, \citenamefont {Wang}, \citenamefont {Bai},\ and\ \citenamefont {Qi}}]{Xu2021_2}%
  \BibitemOpen
  \bibfield  {author} {\bibinfo {author} {\bibfnamefont {L.}~\bibnamefont {Xu}}, \bibinfo {author} {\bibfnamefont {H.}~\bibnamefont {Liu}}, \bibinfo {author} {\bibfnamefont {C.}~\bibnamefont {Song}}, \bibinfo {author} {\bibfnamefont {X.}~\bibnamefont {Li}}, \bibinfo {author} {\bibfnamefont {F.}~\bibnamefont {Li}}, \bibinfo {author} {\bibfnamefont {D.}~\bibnamefont {Li}}, \bibinfo {author} {\bibfnamefont {L.}~\bibnamefont {Wang}}, \bibinfo {author} {\bibfnamefont {X.}~\bibnamefont {Bai}}, \ and\ \bibinfo {author} {\bibfnamefont {J.}~\bibnamefont {Qi}},\ }\href {\doibase 10.1088/2053-1583/abd6b6} {\bibfield  {journal} {\bibinfo  {journal} {2D Materials}\ }\textbf {\bibinfo {volume} {8}} (\bibinfo {year} {2021}{\natexlab{b}}),\ 10.1088/2053-1583/abd6b6}\BibitemShut {NoStop}%
\bibitem [{\citenamefont {F{\"{u}}l{\"{o}}p}\ \emph {et~al.}(2021)\citenamefont {F{\"{u}}l{\"{o}}p}, \citenamefont {M{\'{a}}rffy}, \citenamefont {Zihlmann}, \citenamefont {Gmitra}, \citenamefont {T{\'{o}}v{\'{a}}ri}, \citenamefont {Szentp{\'{e}}teri}, \citenamefont {Kedves}, \citenamefont {Watanabe}, \citenamefont {Taniguchi}, \citenamefont {Fabian}, \citenamefont {Sch{\"{o}}nenberger}, \citenamefont {Makk},\ and\ \citenamefont {Csonka}}]{Fulop2021}%
  \BibitemOpen
  \bibfield  {author} {\bibinfo {author} {\bibfnamefont {B.}~\bibnamefont {F{\"{u}}l{\"{o}}p}}, \bibinfo {author} {\bibfnamefont {A.}~\bibnamefont {M{\'{a}}rffy}}, \bibinfo {author} {\bibfnamefont {S.}~\bibnamefont {Zihlmann}}, \bibinfo {author} {\bibfnamefont {M.}~\bibnamefont {Gmitra}}, \bibinfo {author} {\bibfnamefont {E.}~\bibnamefont {T{\'{o}}v{\'{a}}ri}}, \bibinfo {author} {\bibfnamefont {B.}~\bibnamefont {Szentp{\'{e}}teri}}, \bibinfo {author} {\bibfnamefont {M.}~\bibnamefont {Kedves}}, \bibinfo {author} {\bibfnamefont {K.}~\bibnamefont {Watanabe}}, \bibinfo {author} {\bibfnamefont {T.}~\bibnamefont {Taniguchi}}, \bibinfo {author} {\bibfnamefont {J.}~\bibnamefont {Fabian}}, \bibinfo {author} {\bibfnamefont {C.}~\bibnamefont {Sch{\"{o}}nenberger}}, \bibinfo {author} {\bibfnamefont {P.}~\bibnamefont {Makk}}, \ and\ \bibinfo {author} {\bibfnamefont {S.}~\bibnamefont {Csonka}},\ }\href {\doibase 10.1038/s41699-021-00262-9} {\bibfield  {journal} {\bibinfo  {journal} {npj 2D Materials and Applications}\
  }\textbf {\bibinfo {volume} {5}},\ \bibinfo {pages} {1} (\bibinfo {year} {2021})}\BibitemShut {NoStop}%
\bibitem [{\citenamefont {Kedves}\ \emph {et~al.}(2023)\citenamefont {Kedves}, \citenamefont {Szentp{\'e}teri}, \citenamefont {Márffy}, \citenamefont {Tóvári}, \citenamefont {Papadopoulos}, \citenamefont {Rout}, \citenamefont {Watanabe}, \citenamefont {Taniguchi}, \citenamefont {Goswami}, \citenamefont {Csonka},\ and\ \citenamefont {Makk}}]{Kedves2023}%
  \BibitemOpen
  \bibfield  {author} {\bibinfo {author} {\bibfnamefont {M.}~\bibnamefont {Kedves}}, \bibinfo {author} {\bibfnamefont {B.}~\bibnamefont {Szentp{\'e}teri}}, \bibinfo {author} {\bibfnamefont {A.}~\bibnamefont {Márffy}}, \bibinfo {author} {\bibfnamefont {E.}~\bibnamefont {Tóvári}}, \bibinfo {author} {\bibfnamefont {N.}~\bibnamefont {Papadopoulos}}, \bibinfo {author} {\bibfnamefont {P.~K.}\ \bibnamefont {Rout}}, \bibinfo {author} {\bibfnamefont {K.}~\bibnamefont {Watanabe}}, \bibinfo {author} {\bibfnamefont {T.}~\bibnamefont {Taniguchi}}, \bibinfo {author} {\bibfnamefont {S.}~\bibnamefont {Goswami}}, \bibinfo {author} {\bibfnamefont {S.}~\bibnamefont {Csonka}}, \ and\ \bibinfo {author} {\bibfnamefont {P.}~\bibnamefont {Makk}},\ }\href {\doibase 10.1021/acs.nanolett.3c03029} {\bibfield  {journal} {\bibinfo  {journal} {Nano Letters}\ }\textbf {\bibinfo {volume} {23}},\ \bibinfo {pages} {9508} (\bibinfo {year} {2023})}\BibitemShut {NoStop}%
\bibitem [{\citenamefont {Milivojevi\ifmmode~\acute{c}\else \'{c}\fi{}}\ \emph {et~al.}(2024)\citenamefont {Milivojevi\ifmmode~\acute{c}\else \'{c}\fi{}}, \citenamefont {Kurpas}, \citenamefont {Rassekh}, \citenamefont {Legut},\ and\ \citenamefont {Gmitra}}]{MM2024}%
  \BibitemOpen
  \bibfield  {author} {\bibinfo {author} {\bibfnamefont {M.}~\bibnamefont {Milivojevi\ifmmode~\acute{c}\else \'{c}\fi{}}}, \bibinfo {author} {\bibfnamefont {M.}~\bibnamefont {Kurpas}}, \bibinfo {author} {\bibfnamefont {M.}~\bibnamefont {Rassekh}}, \bibinfo {author} {\bibfnamefont {D.}~\bibnamefont {Legut}}, \ and\ \bibinfo {author} {\bibfnamefont {M.}~\bibnamefont {Gmitra}},\ }\href {\doibase 10.1103/PhysRevB.110.085306} {\bibfield  {journal} {\bibinfo  {journal} {Phys. Rev. B}\ }\textbf {\bibinfo {volume} {110}},\ \bibinfo {pages} {085306} (\bibinfo {year} {2024})}\BibitemShut {NoStop}%
\bibitem [{\citenamefont {Jureczko}, \citenamefont {Milivojević},\ and\ \citenamefont {Kurpas}(2025)}]{Jureczko_2025}%
  \BibitemOpen
  \bibfield  {author} {\bibinfo {author} {\bibfnamefont {P.}~\bibnamefont {Jureczko}}, \bibinfo {author} {\bibfnamefont {M.}~\bibnamefont {Milivojević}}, \ and\ \bibinfo {author} {\bibfnamefont {M.}~\bibnamefont {Kurpas}},\ }\href {\doibase 10.1088/1367-2630/add8b5} {\bibfield  {journal} {\bibinfo  {journal} {New Journal of Physics}\ }\textbf {\bibinfo {volume} {27}},\ \bibinfo {pages} {053006} (\bibinfo {year} {2025})}\BibitemShut {NoStop}%
\bibitem [{\citenamefont {Giannozzi}\ \emph {et~al.}(2009)\citenamefont {Giannozzi}, \citenamefont {Baroni}, \citenamefont {Bonini}, \citenamefont {Calandra}, \citenamefont {Car}, \citenamefont {Cavazzoni}, \citenamefont {Ceresoli}, \citenamefont {Chiarotti}, \citenamefont {Cococcioni}, \citenamefont {Dabo}, \citenamefont {Corso}, \citenamefont {de~Gironcoli}, \citenamefont {Fabris}, \citenamefont {Fratesi}, \citenamefont {Gebauer}, \citenamefont {Gerstmann}, \citenamefont {Gougoussis}, \citenamefont {Kokalj}, \citenamefont {Lazzeri}, \citenamefont {Martin-Samos}, \citenamefont {Marzari}, \citenamefont {Mauri}, \citenamefont {Mazzarello}, \citenamefont {Paolini}, \citenamefont {Pasquarello}, \citenamefont {Paulatto}, \citenamefont {Sbraccia}, \citenamefont {Scandolo}, \citenamefont {Sclauzero}, \citenamefont {Seitsonen}, \citenamefont {Smogunov}, \citenamefont {Umari},\ and\ \citenamefont {Wentzcovitch}}]{Giannozzi_2009}%
  \BibitemOpen
  \bibfield  {author} {\bibinfo {author} {\bibfnamefont {P.}~\bibnamefont {Giannozzi}}, \bibinfo {author} {\bibfnamefont {S.}~\bibnamefont {Baroni}}, \bibinfo {author} {\bibfnamefont {N.}~\bibnamefont {Bonini}}, \bibinfo {author} {\bibfnamefont {M.}~\bibnamefont {Calandra}}, \bibinfo {author} {\bibfnamefont {R.}~\bibnamefont {Car}}, \bibinfo {author} {\bibfnamefont {C.}~\bibnamefont {Cavazzoni}}, \bibinfo {author} {\bibfnamefont {D.}~\bibnamefont {Ceresoli}}, \bibinfo {author} {\bibfnamefont {G.~L.}\ \bibnamefont {Chiarotti}}, \bibinfo {author} {\bibfnamefont {M.}~\bibnamefont {Cococcioni}}, \bibinfo {author} {\bibfnamefont {I.}~\bibnamefont {Dabo}}, \bibinfo {author} {\bibfnamefont {A.~D.}\ \bibnamefont {Corso}}, \bibinfo {author} {\bibfnamefont {S.}~\bibnamefont {de~Gironcoli}}, \bibinfo {author} {\bibfnamefont {S.}~\bibnamefont {Fabris}}, \bibinfo {author} {\bibfnamefont {G.}~\bibnamefont {Fratesi}}, \bibinfo {author} {\bibfnamefont {R.}~\bibnamefont {Gebauer}}, \bibinfo {author} {\bibfnamefont
  {U.}~\bibnamefont {Gerstmann}}, \bibinfo {author} {\bibfnamefont {C.}~\bibnamefont {Gougoussis}}, \bibinfo {author} {\bibfnamefont {A.}~\bibnamefont {Kokalj}}, \bibinfo {author} {\bibfnamefont {M.}~\bibnamefont {Lazzeri}}, \bibinfo {author} {\bibfnamefont {L.}~\bibnamefont {Martin-Samos}}, \bibinfo {author} {\bibfnamefont {N.}~\bibnamefont {Marzari}}, \bibinfo {author} {\bibfnamefont {F.}~\bibnamefont {Mauri}}, \bibinfo {author} {\bibfnamefont {R.}~\bibnamefont {Mazzarello}}, \bibinfo {author} {\bibfnamefont {S.}~\bibnamefont {Paolini}}, \bibinfo {author} {\bibfnamefont {A.}~\bibnamefont {Pasquarello}}, \bibinfo {author} {\bibfnamefont {L.}~\bibnamefont {Paulatto}}, \bibinfo {author} {\bibfnamefont {C.}~\bibnamefont {Sbraccia}}, \bibinfo {author} {\bibfnamefont {S.}~\bibnamefont {Scandolo}}, \bibinfo {author} {\bibfnamefont {G.}~\bibnamefont {Sclauzero}}, \bibinfo {author} {\bibfnamefont {A.~P.}\ \bibnamefont {Seitsonen}}, \bibinfo {author} {\bibfnamefont {A.}~\bibnamefont {Smogunov}}, \bibinfo {author}
  {\bibfnamefont {P.}~\bibnamefont {Umari}}, \ and\ \bibinfo {author} {\bibfnamefont {R.~M.}\ \bibnamefont {Wentzcovitch}},\ }\href {\doibase 10.1088/0953-8984/21/39/395502} {\bibfield  {journal} {\bibinfo  {journal} {Journal of Physics: Condensed Matter}\ }\textbf {\bibinfo {volume} {21}},\ \bibinfo {pages} {395502} (\bibinfo {year} {2009})}\BibitemShut {NoStop}%
\bibitem [{\citenamefont {Giannozzi}\ \emph {et~al.}(2017)\citenamefont {Giannozzi}, \citenamefont {Andreussi}, \citenamefont {Brumme}, \citenamefont {Bunau}, \citenamefont {Nardelli}, \citenamefont {Calandra}, \citenamefont {Car}, \citenamefont {Cavazzoni}, \citenamefont {Ceresoli}, \citenamefont {Cococcioni}, \citenamefont {Colonna}, \citenamefont {Carnimeo}, \citenamefont {Corso}, \citenamefont {de~Gironcoli}, \citenamefont {Delugas}, \citenamefont {DiStasio}, \citenamefont {Ferretti}, \citenamefont {Floris}, \citenamefont {Fratesi}, \citenamefont {Fugallo}, \citenamefont {Gebauer}, \citenamefont {Gerstmann}, \citenamefont {Giustino}, \citenamefont {Gorni}, \citenamefont {Jia}, \citenamefont {Kawamura}, \citenamefont {Ko}, \citenamefont {Kokalj}, \citenamefont {Küçükbenli}, \citenamefont {Lazzeri}, \citenamefont {Marsili}, \citenamefont {Marzari}, \citenamefont {Mauri}, \citenamefont {Nguyen}, \citenamefont {Nguyen}, \citenamefont {de-la Roza}, \citenamefont {Paulatto}, \citenamefont {Poncé},
  \citenamefont {Rocca}, \citenamefont {Sabatini}, \citenamefont {Santra}, \citenamefont {Schlipf}, \citenamefont {Seitsonen}, \citenamefont {Smogunov}, \citenamefont {Timrov}, \citenamefont {Thonhauser}, \citenamefont {Umari}, \citenamefont {Vast}, \citenamefont {Wu},\ and\ \citenamefont {Baroni}}]{Giannozzi_2017}%
  \BibitemOpen
  \bibfield  {author} {\bibinfo {author} {\bibfnamefont {P.}~\bibnamefont {Giannozzi}}, \bibinfo {author} {\bibfnamefont {O.}~\bibnamefont {Andreussi}}, \bibinfo {author} {\bibfnamefont {T.}~\bibnamefont {Brumme}}, \bibinfo {author} {\bibfnamefont {O.}~\bibnamefont {Bunau}}, \bibinfo {author} {\bibfnamefont {M.~B.}\ \bibnamefont {Nardelli}}, \bibinfo {author} {\bibfnamefont {M.}~\bibnamefont {Calandra}}, \bibinfo {author} {\bibfnamefont {R.}~\bibnamefont {Car}}, \bibinfo {author} {\bibfnamefont {C.}~\bibnamefont {Cavazzoni}}, \bibinfo {author} {\bibfnamefont {D.}~\bibnamefont {Ceresoli}}, \bibinfo {author} {\bibfnamefont {M.}~\bibnamefont {Cococcioni}}, \bibinfo {author} {\bibfnamefont {N.}~\bibnamefont {Colonna}}, \bibinfo {author} {\bibfnamefont {I.}~\bibnamefont {Carnimeo}}, \bibinfo {author} {\bibfnamefont {A.~D.}\ \bibnamefont {Corso}}, \bibinfo {author} {\bibfnamefont {S.}~\bibnamefont {de~Gironcoli}}, \bibinfo {author} {\bibfnamefont {P.}~\bibnamefont {Delugas}}, \bibinfo {author} {\bibfnamefont
  {R.~A.}\ \bibnamefont {DiStasio}}, \bibinfo {author} {\bibfnamefont {A.}~\bibnamefont {Ferretti}}, \bibinfo {author} {\bibfnamefont {A.}~\bibnamefont {Floris}}, \bibinfo {author} {\bibfnamefont {G.}~\bibnamefont {Fratesi}}, \bibinfo {author} {\bibfnamefont {G.}~\bibnamefont {Fugallo}}, \bibinfo {author} {\bibfnamefont {R.}~\bibnamefont {Gebauer}}, \bibinfo {author} {\bibfnamefont {U.}~\bibnamefont {Gerstmann}}, \bibinfo {author} {\bibfnamefont {F.}~\bibnamefont {Giustino}}, \bibinfo {author} {\bibfnamefont {T.}~\bibnamefont {Gorni}}, \bibinfo {author} {\bibfnamefont {J.}~\bibnamefont {Jia}}, \bibinfo {author} {\bibfnamefont {M.}~\bibnamefont {Kawamura}}, \bibinfo {author} {\bibfnamefont {H.-Y.}\ \bibnamefont {Ko}}, \bibinfo {author} {\bibfnamefont {A.}~\bibnamefont {Kokalj}}, \bibinfo {author} {\bibfnamefont {E.}~\bibnamefont {Küçükbenli}}, \bibinfo {author} {\bibfnamefont {M.}~\bibnamefont {Lazzeri}}, \bibinfo {author} {\bibfnamefont {M.}~\bibnamefont {Marsili}}, \bibinfo {author} {\bibfnamefont
  {N.}~\bibnamefont {Marzari}}, \bibinfo {author} {\bibfnamefont {F.}~\bibnamefont {Mauri}}, \bibinfo {author} {\bibfnamefont {N.~L.}\ \bibnamefont {Nguyen}}, \bibinfo {author} {\bibfnamefont {H.-V.}\ \bibnamefont {Nguyen}}, \bibinfo {author} {\bibfnamefont {A.~O.}\ \bibnamefont {de-la Roza}}, \bibinfo {author} {\bibfnamefont {L.}~\bibnamefont {Paulatto}}, \bibinfo {author} {\bibfnamefont {S.}~\bibnamefont {Poncé}}, \bibinfo {author} {\bibfnamefont {D.}~\bibnamefont {Rocca}}, \bibinfo {author} {\bibfnamefont {R.}~\bibnamefont {Sabatini}}, \bibinfo {author} {\bibfnamefont {B.}~\bibnamefont {Santra}}, \bibinfo {author} {\bibfnamefont {M.}~\bibnamefont {Schlipf}}, \bibinfo {author} {\bibfnamefont {A.~P.}\ \bibnamefont {Seitsonen}}, \bibinfo {author} {\bibfnamefont {A.}~\bibnamefont {Smogunov}}, \bibinfo {author} {\bibfnamefont {I.}~\bibnamefont {Timrov}}, \bibinfo {author} {\bibfnamefont {T.}~\bibnamefont {Thonhauser}}, \bibinfo {author} {\bibfnamefont {P.}~\bibnamefont {Umari}}, \bibinfo {author}
  {\bibfnamefont {N.}~\bibnamefont {Vast}}, \bibinfo {author} {\bibfnamefont {X.}~\bibnamefont {Wu}}, \ and\ \bibinfo {author} {\bibfnamefont {S.}~\bibnamefont {Baroni}},\ }\href {\doibase 10.1088/1361-648X/aa8f79} {\bibfield  {journal} {\bibinfo  {journal} {Journal of Physics: Condensed Matter}\ }\textbf {\bibinfo {volume} {29}},\ \bibinfo {pages} {465901} (\bibinfo {year} {2017})}\BibitemShut {NoStop}%
\bibitem [{\citenamefont {Hamann}(2013)}]{hamnann_oncv}%
  \BibitemOpen
  \bibfield  {author} {\bibinfo {author} {\bibfnamefont {D.~R.}\ \bibnamefont {Hamann}},\ }\href {\doibase 10.1103/PhysRevB.88.085117} {\bibfield  {journal} {\bibinfo  {journal} {Phys. Rev. B}\ }\textbf {\bibinfo {volume} {88}},\ \bibinfo {pages} {085117} (\bibinfo {year} {2013})}\BibitemShut {NoStop}%
\bibitem [{\citenamefont {Perdew}, \citenamefont {Burke},\ and\ \citenamefont {Ernzerhof}(1996)}]{PBE}%
  \BibitemOpen
  \bibfield  {author} {\bibinfo {author} {\bibfnamefont {J.~P.}\ \bibnamefont {Perdew}}, \bibinfo {author} {\bibfnamefont {K.}~\bibnamefont {Burke}}, \ and\ \bibinfo {author} {\bibfnamefont {M.}~\bibnamefont {Ernzerhof}},\ }\href {\doibase 10.1103/PhysRevLett.77.3865} {\bibfield  {journal} {\bibinfo  {journal} {Phys. Rev. Lett.}\ }\textbf {\bibinfo {volume} {77}},\ \bibinfo {pages} {3865} (\bibinfo {year} {1996})}\BibitemShut {NoStop}%
\bibitem [{\citenamefont {Heyd}, \citenamefont {Scuseria},\ and\ \citenamefont {Ernzerhof}(2003)}]{HSE}%
  \BibitemOpen
  \bibfield  {author} {\bibinfo {author} {\bibfnamefont {J.}~\bibnamefont {Heyd}}, \bibinfo {author} {\bibfnamefont {G.~E.}\ \bibnamefont {Scuseria}}, \ and\ \bibinfo {author} {\bibfnamefont {M.}~\bibnamefont {Ernzerhof}},\ }\href {\doibase 10.1063/1.1564060} {\bibfield  {journal} {\bibinfo  {journal} {The Journal of Chemical Physics}\ }\textbf {\bibinfo {volume} {118}},\ \bibinfo {pages} {8207} (\bibinfo {year} {2003})}\BibitemShut {NoStop}%
\bibitem [{\citenamefont {Uchida}\ \emph {et~al.}(2014)\citenamefont {Uchida}, \citenamefont {Furuya}, \citenamefont {Iwata},\ and\ \citenamefont {Oshiyama}}]{Uchida2014}%
  \BibitemOpen
  \bibfield  {author} {\bibinfo {author} {\bibfnamefont {K.}~\bibnamefont {Uchida}}, \bibinfo {author} {\bibfnamefont {S.}~\bibnamefont {Furuya}}, \bibinfo {author} {\bibfnamefont {J.-I.}\ \bibnamefont {Iwata}}, \ and\ \bibinfo {author} {\bibfnamefont {A.}~\bibnamefont {Oshiyama}},\ }\href {\doibase 10.1103/PhysRevB.90.155451} {\bibfield  {journal} {\bibinfo  {journal} {Phys. Rev. B}\ }\textbf {\bibinfo {volume} {90}},\ \bibinfo {pages} {155451} (\bibinfo {year} {2014})}\BibitemShut {NoStop}%
\bibitem [{\citenamefont {Ambrosch-Draxl}\ and\ \citenamefont {Sofo}(2006)}]{Ambrosch-Draxl2006}%
  \BibitemOpen
  \bibfield  {author} {\bibinfo {author} {\bibfnamefont {C.}~\bibnamefont {Ambrosch-Draxl}}\ and\ \bibinfo {author} {\bibfnamefont {J.~O.}\ \bibnamefont {Sofo}},\ }\href {\doibase 10.1016/j.cpc.2006.03.005} {\bibfield  {journal} {\bibinfo  {journal} {Computer Physics Communications}\ }\textbf {\bibinfo {volume} {175}},\ \bibinfo {pages} {1} (\bibinfo {year} {2006})}\BibitemShut {NoStop}%
\bibitem [{\citenamefont {Adler}(1962)}]{Adler1962}%
  \BibitemOpen
  \bibfield  {author} {\bibinfo {author} {\bibfnamefont {S.~L.}\ \bibnamefont {Adler}},\ }\href {\doibase 10.1103/PhysRev.126.413} {\bibfield  {journal} {\bibinfo  {journal} {Physical Review}\ }\textbf {\bibinfo {volume} {126}},\ \bibinfo {pages} {413} (\bibinfo {year} {1962})}\BibitemShut {NoStop}%
\bibitem [{\citenamefont {Kurpas}\ and\ \citenamefont {Fabian}(2021)}]{Kurpas2021}%
  \BibitemOpen
  \bibfield  {author} {\bibinfo {author} {\bibfnamefont {M.}~\bibnamefont {Kurpas}}\ and\ \bibinfo {author} {\bibfnamefont {J.}~\bibnamefont {Fabian}},\ }\href {\doibase 10.1103/PhysRevB.103.125409} {\bibfield  {journal} {\bibinfo  {journal} {Physical Review B}\ }\textbf {\bibinfo {volume} {103}},\ \bibinfo {pages} {1} (\bibinfo {year} {2021})}\BibitemShut {NoStop}%
\bibitem [{\citenamefont {Kandemir}\ \emph {et~al.}(2018)\citenamefont {Kandemir}, \citenamefont {Akbali}, \citenamefont {Kahraman}, \citenamefont {Badalov}, \citenamefont {Ozcan}, \citenamefont {Iyikanat},\ and\ \citenamefont {Sahin}}]{Kandemir2018}%
  \BibitemOpen
  \bibfield  {author} {\bibinfo {author} {\bibfnamefont {A.}~\bibnamefont {Kandemir}}, \bibinfo {author} {\bibfnamefont {B.}~\bibnamefont {Akbali}}, \bibinfo {author} {\bibfnamefont {Z.}~\bibnamefont {Kahraman}}, \bibinfo {author} {\bibfnamefont {S.~V.}\ \bibnamefont {Badalov}}, \bibinfo {author} {\bibfnamefont {M.}~\bibnamefont {Ozcan}}, \bibinfo {author} {\bibfnamefont {F.}~\bibnamefont {Iyikanat}}, \ and\ \bibinfo {author} {\bibfnamefont {H.}~\bibnamefont {Sahin}},\ }\href {\doibase 10.1088/1361-6641/aacba2} {\bibfield  {journal} {\bibinfo  {journal} {Semiconductor Science and Technology}\ }\textbf {\bibinfo {volume} {33}} (\bibinfo {year} {2018}),\ 10.1088/1361-6641/aacba2}\BibitemShut {NoStop}%
\bibitem [{\citenamefont {Liu}\ \emph {et~al.}(2014)\citenamefont {Liu}, \citenamefont {Zhang}, \citenamefont {Cao}, \citenamefont {Jin}, \citenamefont {Qiu}, \citenamefont {Zhou}, \citenamefont {Zettl}, \citenamefont {Yang}, \citenamefont {Louie},\ and\ \citenamefont {Wang}}]{Liu2014_vdw}%
  \BibitemOpen
  \bibfield  {author} {\bibinfo {author} {\bibfnamefont {K.}~\bibnamefont {Liu}}, \bibinfo {author} {\bibfnamefont {L.}~\bibnamefont {Zhang}}, \bibinfo {author} {\bibfnamefont {T.}~\bibnamefont {Cao}}, \bibinfo {author} {\bibfnamefont {C.}~\bibnamefont {Jin}}, \bibinfo {author} {\bibfnamefont {D.}~\bibnamefont {Qiu}}, \bibinfo {author} {\bibfnamefont {Q.}~\bibnamefont {Zhou}}, \bibinfo {author} {\bibfnamefont {A.}~\bibnamefont {Zettl}}, \bibinfo {author} {\bibfnamefont {P.}~\bibnamefont {Yang}}, \bibinfo {author} {\bibfnamefont {S.~G.}\ \bibnamefont {Louie}}, \ and\ \bibinfo {author} {\bibfnamefont {F.}~\bibnamefont {Wang}},\ }\href {\doibase 10.1038/ncomms5966} {\bibfield  {journal} {\bibinfo  {journal} {Nature Communications}\ }\textbf {\bibinfo {volume} {5}},\ \bibinfo {pages} {4966} (\bibinfo {year} {2014})}\BibitemShut {NoStop}%
\bibitem [{\citenamefont {Nayak}\ \emph {et~al.}(2017)\citenamefont {Nayak}, \citenamefont {Horbatenko}, \citenamefont {Ahn}, \citenamefont {Kim}, \citenamefont {Lee}, \citenamefont {Ma}, \citenamefont {Jang}, \citenamefont {Lim}, \citenamefont {Kim}, \citenamefont {Ryu}, \citenamefont {Cheong}, \citenamefont {Park},\ and\ \citenamefont {Shin}}]{Nayak2017}%
  \BibitemOpen
  \bibfield  {author} {\bibinfo {author} {\bibfnamefont {P.~K.}\ \bibnamefont {Nayak}}, \bibinfo {author} {\bibfnamefont {Y.}~\bibnamefont {Horbatenko}}, \bibinfo {author} {\bibfnamefont {S.}~\bibnamefont {Ahn}}, \bibinfo {author} {\bibfnamefont {G.}~\bibnamefont {Kim}}, \bibinfo {author} {\bibfnamefont {J.-U.}\ \bibnamefont {Lee}}, \bibinfo {author} {\bibfnamefont {K.~Y.}\ \bibnamefont {Ma}}, \bibinfo {author} {\bibfnamefont {A.-R.}\ \bibnamefont {Jang}}, \bibinfo {author} {\bibfnamefont {H.}~\bibnamefont {Lim}}, \bibinfo {author} {\bibfnamefont {D.}~\bibnamefont {Kim}}, \bibinfo {author} {\bibfnamefont {S.}~\bibnamefont {Ryu}}, \bibinfo {author} {\bibfnamefont {H.}~\bibnamefont {Cheong}}, \bibinfo {author} {\bibfnamefont {N.}~\bibnamefont {Park}}, \ and\ \bibinfo {author} {\bibfnamefont {H.~S.}\ \bibnamefont {Shin}},\ }\href {\doibase 10.1021/acsnano.7b00640} {\bibfield  {journal} {\bibinfo  {journal} {ACS Nano}\ }\textbf {\bibinfo {volume} {11}},\ \bibinfo {pages} {4041} (\bibinfo {year} {2017})},\
  \bibinfo {note} {pMID: 28363013}\BibitemShut {NoStop}%
\bibitem [{\citenamefont {Yan}\ \emph {et~al.}(2019)\citenamefont {Yan}, \citenamefont {Meng}, \citenamefont {Meng}, \citenamefont {Weng}, \citenamefont {Kang},\ and\ \citenamefont {Li}}]{Yan2019}%
  \BibitemOpen
  \bibfield  {author} {\bibinfo {author} {\bibfnamefont {W.}~\bibnamefont {Yan}}, \bibinfo {author} {\bibfnamefont {L.}~\bibnamefont {Meng}}, \bibinfo {author} {\bibfnamefont {Z.}~\bibnamefont {Meng}}, \bibinfo {author} {\bibfnamefont {Y.}~\bibnamefont {Weng}}, \bibinfo {author} {\bibfnamefont {L.}~\bibnamefont {Kang}}, \ and\ \bibinfo {author} {\bibfnamefont {X.-a.}\ \bibnamefont {Li}},\ }\href {\doibase 10.1021/acs.jpcc.9b08602} {\bibfield  {journal} {\bibinfo  {journal} {The Journal of Physical Chemistry C}\ }\textbf {\bibinfo {volume} {123}},\ \bibinfo {pages} {30684} (\bibinfo {year} {2019})}\BibitemShut {NoStop}%
\bibitem [{\citenamefont {Ahn}\ \emph {et~al.}(2024)\citenamefont {Ahn}, \citenamefont {Kang}, \citenamefont {Yoon},\ and\ \citenamefont {Krogel}}]{Ahn2024}%
  \BibitemOpen
  \bibfield  {author} {\bibinfo {author} {\bibfnamefont {J.}~\bibnamefont {Ahn}}, \bibinfo {author} {\bibfnamefont {S.-H.}\ \bibnamefont {Kang}}, \bibinfo {author} {\bibfnamefont {M.}~\bibnamefont {Yoon}}, \ and\ \bibinfo {author} {\bibfnamefont {J.~T.}\ \bibnamefont {Krogel}},\ }\href {\doibase 10.1103/PhysRevResearch.6.033177} {\bibfield  {journal} {\bibinfo  {journal} {Phys. Rev. Res.}\ }\textbf {\bibinfo {volume} {6}},\ \bibinfo {pages} {033177} (\bibinfo {year} {2024})}\BibitemShut {NoStop}%
\bibitem [{\citenamefont {Tharrault}\ \emph {et~al.}(2025)\citenamefont {Tharrault}, \citenamefont {Ayari}, \citenamefont {Arfaoui}, \citenamefont {Desgué}, \citenamefont {Goff}, \citenamefont {Morfin}, \citenamefont {Palomo}, \citenamefont {Rosticher}, \citenamefont {Jaziri}, \citenamefont {Plaçais}, \citenamefont {Legagneux}, \citenamefont {Carosella}, \citenamefont {Voisin}, \citenamefont {Ferreira},\ and\ \citenamefont {Baudin}}]{Tharrault2025}%
  \BibitemOpen
  \bibfield  {author} {\bibinfo {author} {\bibfnamefont {M.}~\bibnamefont {Tharrault}}, \bibinfo {author} {\bibfnamefont {S.}~\bibnamefont {Ayari}}, \bibinfo {author} {\bibfnamefont {M.}~\bibnamefont {Arfaoui}}, \bibinfo {author} {\bibfnamefont {E.}~\bibnamefont {Desgué}}, \bibinfo {author} {\bibfnamefont {R.~L.}\ \bibnamefont {Goff}}, \bibinfo {author} {\bibfnamefont {P.}~\bibnamefont {Morfin}}, \bibinfo {author} {\bibfnamefont {J.}~\bibnamefont {Palomo}}, \bibinfo {author} {\bibfnamefont {M.}~\bibnamefont {Rosticher}}, \bibinfo {author} {\bibfnamefont {S.}~\bibnamefont {Jaziri}}, \bibinfo {author} {\bibfnamefont {B.}~\bibnamefont {Plaçais}}, \bibinfo {author} {\bibfnamefont {P.}~\bibnamefont {Legagneux}}, \bibinfo {author} {\bibfnamefont {F.}~\bibnamefont {Carosella}}, \bibinfo {author} {\bibfnamefont {C.}~\bibnamefont {Voisin}}, \bibinfo {author} {\bibfnamefont {R.}~\bibnamefont {Ferreira}}, \ and\ \bibinfo {author} {\bibfnamefont {E.}~\bibnamefont {Baudin}},\ }\href {\doibase
  10.1103/PhysRevLett.134.066901} {\bibfield  {journal} {\bibinfo  {journal} {Physical Review Letters}\ }\textbf {\bibinfo {volume} {134}},\ \bibinfo {pages} {066901} (\bibinfo {year} {2025})}\BibitemShut {NoStop}%
\end{thebibliography}%

\onecolumngrid
\section*{Supplemental Material: Optimizing optical properties of bilayer PtSe$_2$: the role of twist angle and hydrostatic pressure}
\subsection*{Structural and electronic properties of PtSe$_2$ bilayers under hydrostatic pressure }
\begin{figure*}[h]
    \centering
\includegraphics[width=0.7\textwidth]{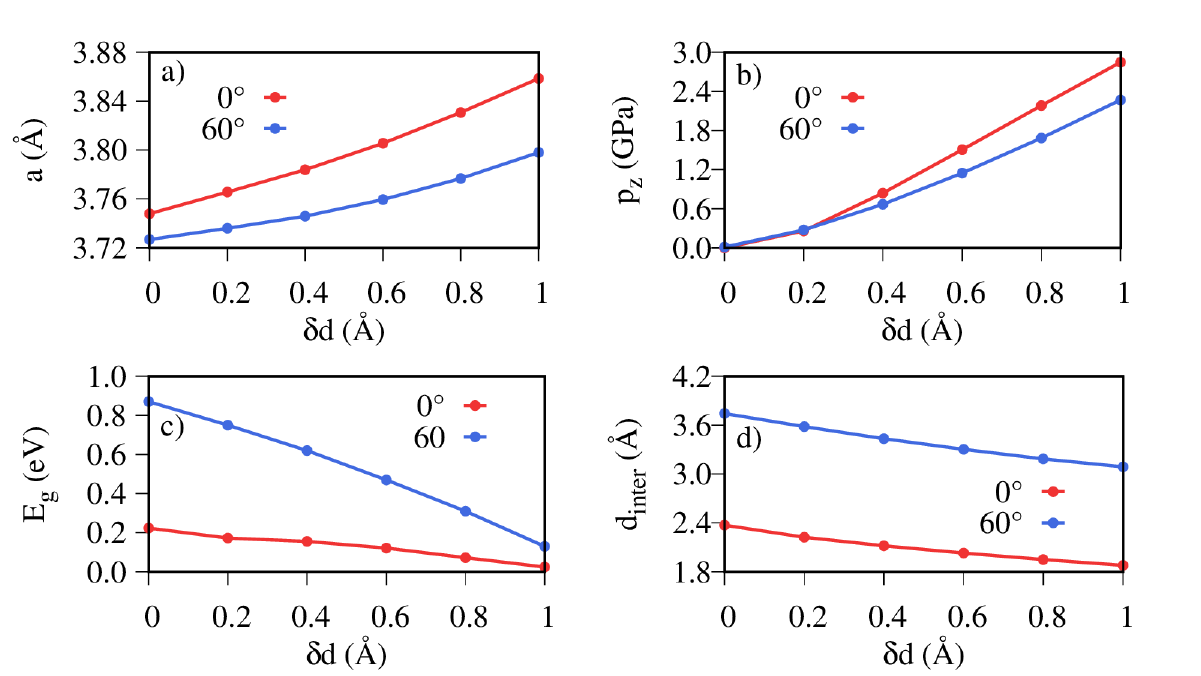}
    \caption{Parameters of strained 0$^\circ$ and 60$^\circ$ systems; ~a)\,cell parameter,~b)\,pressure along z-direction,~c)\,size of the band gap,~d)\,interlayer distance along z-direction\,(between selenium atoms).}
    \label{fig:parameters}
\end{figure*}

In Fig. \ref{fig:parameters} we show  how the structural and electronic properties of  AA and AB bilayers change with increasing hydrostatic pressure defined by $\delta$d (thickness reduction of the bilayer). We see, Fig. \ref{fig:parameters}  a), that with growing $\delta$d the lattice constant expands to relax the generated in-plane stress. The corresponding out-of-plane pressure  and the reduction of the interlayer distance $d_{\rm{inter}}$ are presented in Fig. \ref{fig:parameters}  a) and c), respectively.  The relative structural response of AA and AB bilayers to the applied pressure is similar, however, a slightly bigger pressure values for  AA than for AB case are observed, most probably due to the much smaller initial interlayer distance.

\subsection*{Charge density profile}

\begin{figure*}
    \centering
    \includegraphics[width = 0.8\columnwidth]{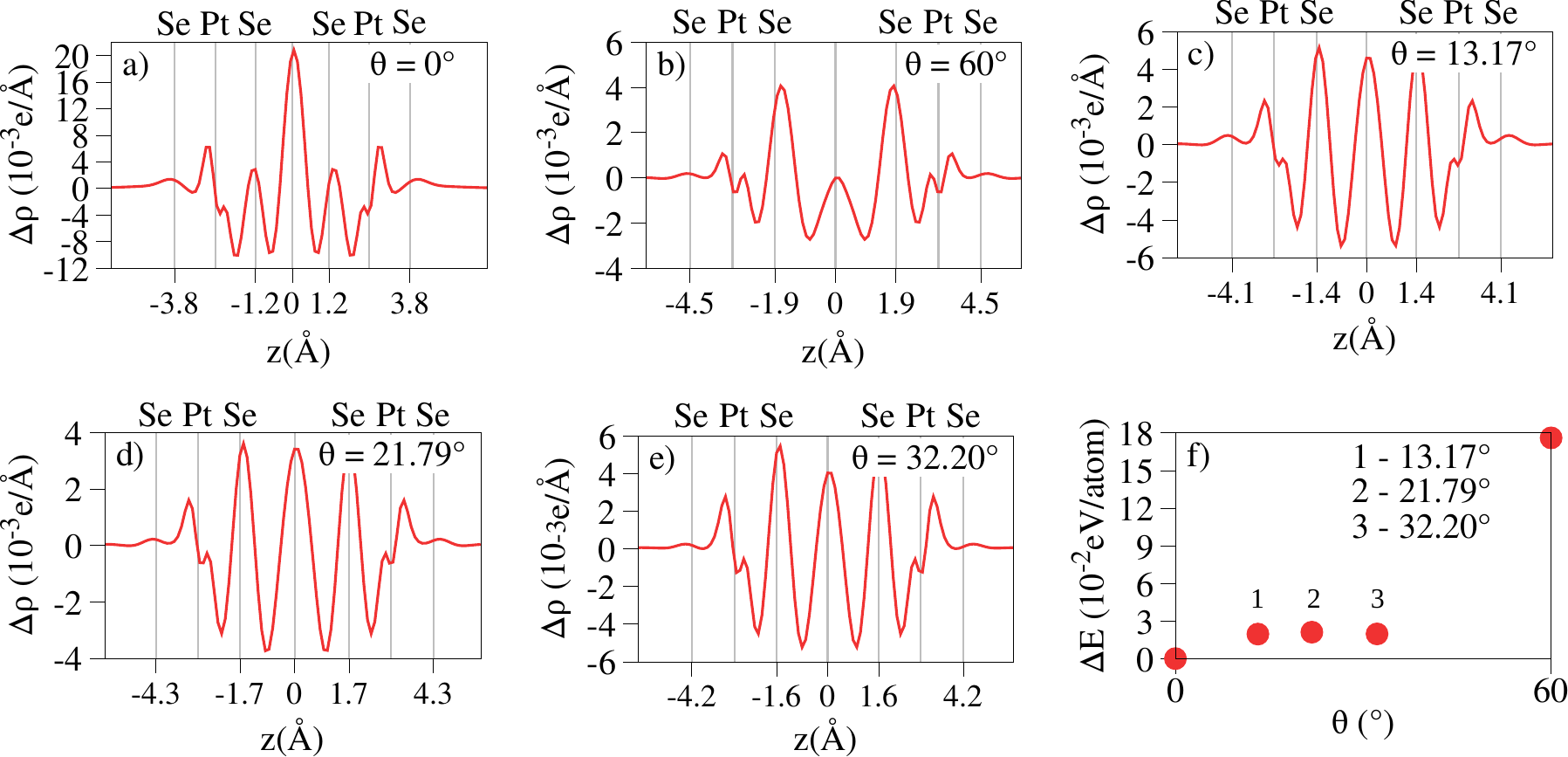} 
    \caption{Charge density difference profile of the a)~0$^\circ$, b)~60$^\circ$, c)~13.17$^\circ$, d)~21.79$^\circ$, e)~32.20$^\circ$ and f)~comparison of total energy difference for all systems, with regard to AA.   \label{fig:cden}}
   
\end{figure*}

To get an insight into the details of  the interlayer interaction, we calculate the charge density profile $\Delta \rho$ within the simulation cell for each bilayer system. We define $\Delta \rho$ as  the difference of charge density of the bilayer minus charge density of two non-interacting monolayers placed at the same positions as in the bilayer:
\begin{equation}
    \Delta\rho(z) = \int_{S} \left(\rho_{\rm{BL}} - [\rho_{\rm{U}} + \rho_{\rm{L}}]\right) dxy,
    \label{eq:delta_rho}
\end{equation}
where $\rho_{BL}$ is the charge density of the bilayer, and $\rho_{U}$ and $\rho_{L}$ are charge densities of the upper and lower monolayer, respectively, and $S$ is the area of the unit cell in the $xy$ plane. 
Qualitative differences between the AA ($\theta=0^\circ$), and twisted bilayers are evident, see Fig. \ref{fig:cden}. 

For comparison, the band structure of single-layer PtSe$_2$ is shown in Fig.~\ref{fig_bnd_1} a).

\subsection*{Comparison of non-relativistic and relativistic calculations}
In heavy element materials, spin-orbit coupling considerably modifies the band structure, opening spin-orbital gaps of hundreds meV. In PtSe$_2$ SOC opens orbital gaps from tens meV up to 300\,meV at the $\Gamma$ point in the valence band \cite{Kurpas2021}, thus it cannot be neglected. 

\begin{figure*}[hb]
    \centering
   \includegraphics[width=0.85\textwidth]{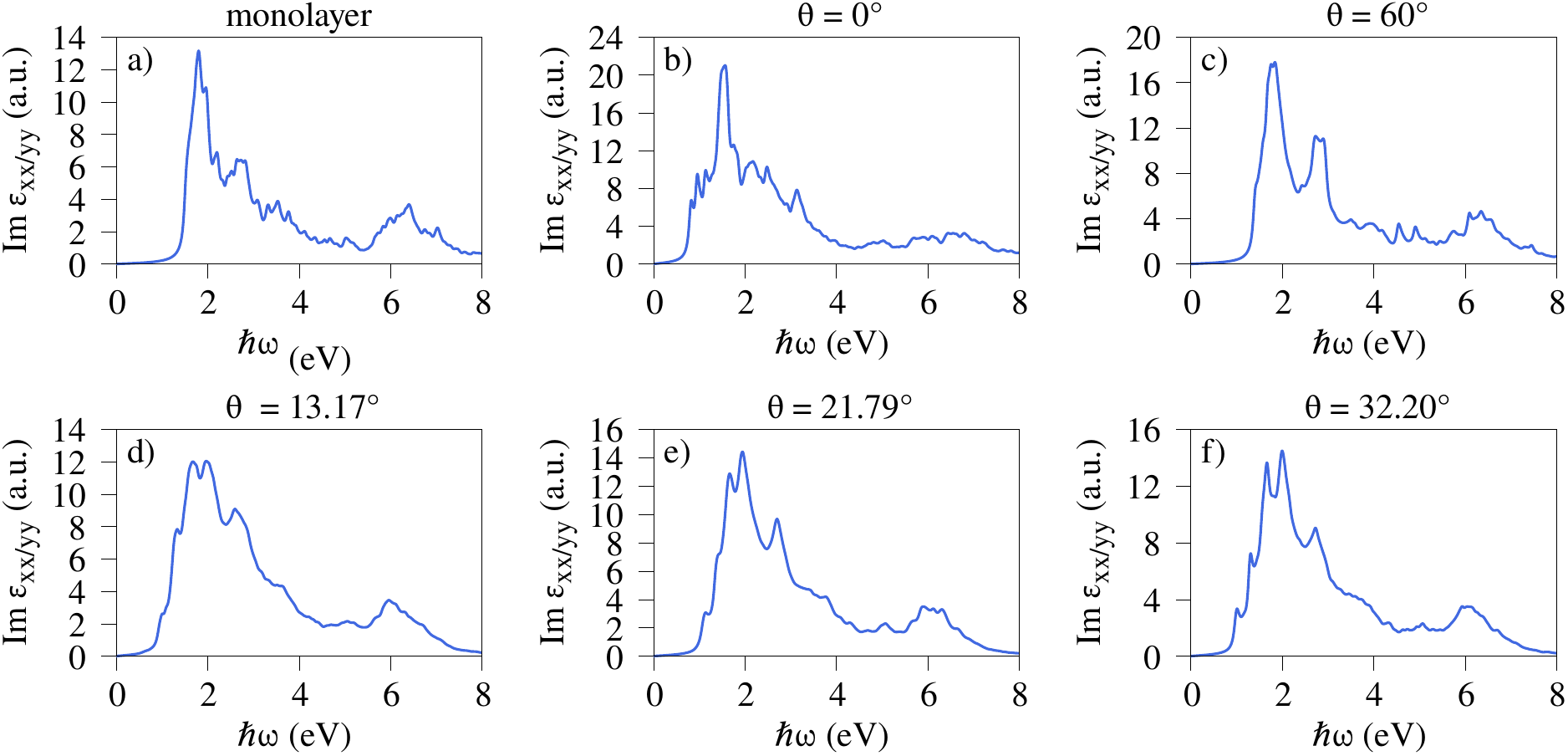}
    \caption{Imaginary part of dielectric tensor for a) monolayer, b) 0$^\circ$, c)\,60$^{\circ}$, d)\,13.17$^\circ$, e)\,21.79$^\circ$ and 
    f)\,32.20$^{\circ}$ calculated without spin-orbit coupling interaction. }     
    \label{figs:opt_nrel}
\end{figure*}

In Fig. \ref{figs:opt_nrel} we show the calculated imaginary part of the dielectric function. A direct comparison with Fig. 2 in the main text indicates, that  \textit{Im}$\epsilon_{xx/yy}$ for relativistic case displays slightly smoother shape than for non-relativistic one due lifted band degeneracy, while the overall shape of \textit{Im}$\epsilon_{xx/yy}$ remains the same. 

\begin{figure*}
    \centering
    \includegraphics[width = 0.99\linewidth]{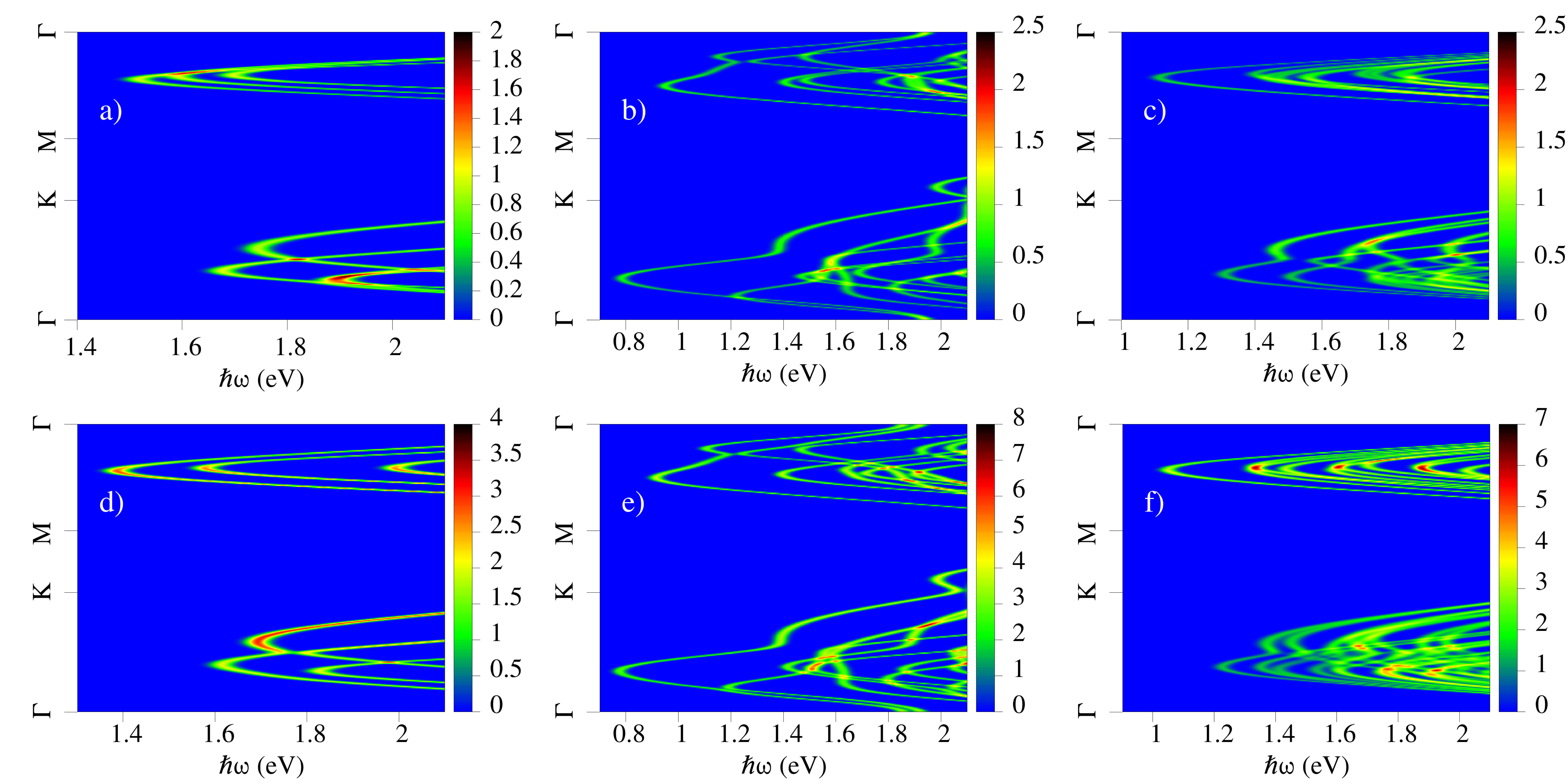}
    \caption{\label{fig:mtx_elements} Comparison of momentum matrix elements calculated without (top row) and with spin-orbit coupling (bottom row) for the monolayer (left column), AA bilayer (middle column) and AB bilayer  PtSe$_2$ (right column). For relativistic calculation in monolayer and  AA case an infinitesimal ($10^{-6}$\,V/nm) transverse electric field was applied to remove the bands degeneracy and the arbitrariness of choice of eigenvectors. }
\end{figure*}

This is confirmed by looking at momentum matrix elements calculated using non-relativistic and relativistic approximations shown in Fig. \ref{fig:mtx_elements}. As can be seen, the inclusion of SOC does not lead to qualitatively different results. Thus, the non-relativistic approximation is valid for explaining essential features of optical transitions.

\begin{figure*}
    \centering
    \includegraphics[width=0.99\linewidth]{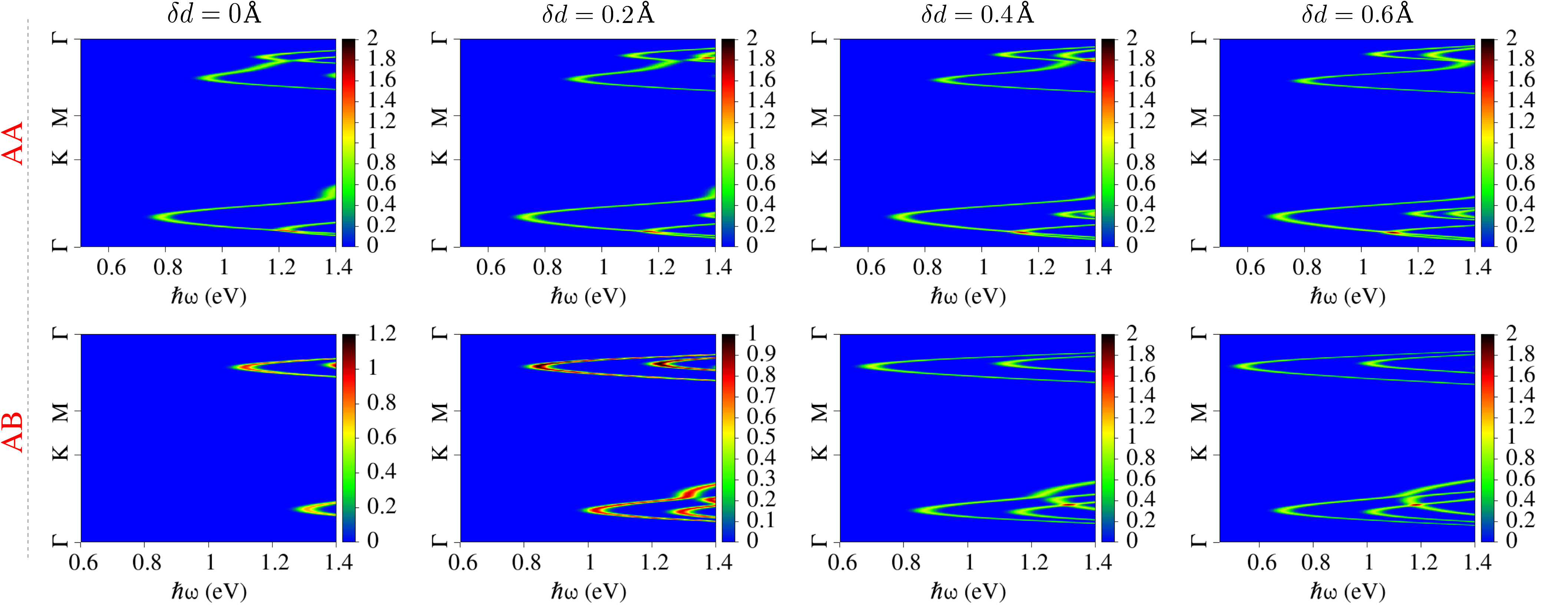}
    \caption{Momentum matrix elements for the AA (top row) and AB (bottom row) calculated  for k-points along high symmetry lines of the first Brillouin zone  versus photon energy $\hbar \omega$  and for several values of hydrostatic pressure. Calculation without spin-orbit coupling.  }
    \label{fig:enter-label}
\end{figure*}

\newpage
\subsection*{Comparison of PBE and hybrid exchange-correlation functionals}
We have performed first principles calculations using the hybrid Heyd-Scuseria-Ernzerhof\,(HSE) functional, assuming  20\%  contribution of the Fock exchange \cite{HSE}. The calculated band structures shown in Fig. \ref{fig_bnd_1} a) - c) show minor discrepancies between PBE and HSE functionals, except a constant energy offset of the conduction bands. Similar conclusions were drawn in Ref. \cite{Tharrault2025}. 

\begin{figure*}[h]
    \centering
    \includegraphics[width=0.84\textwidth]{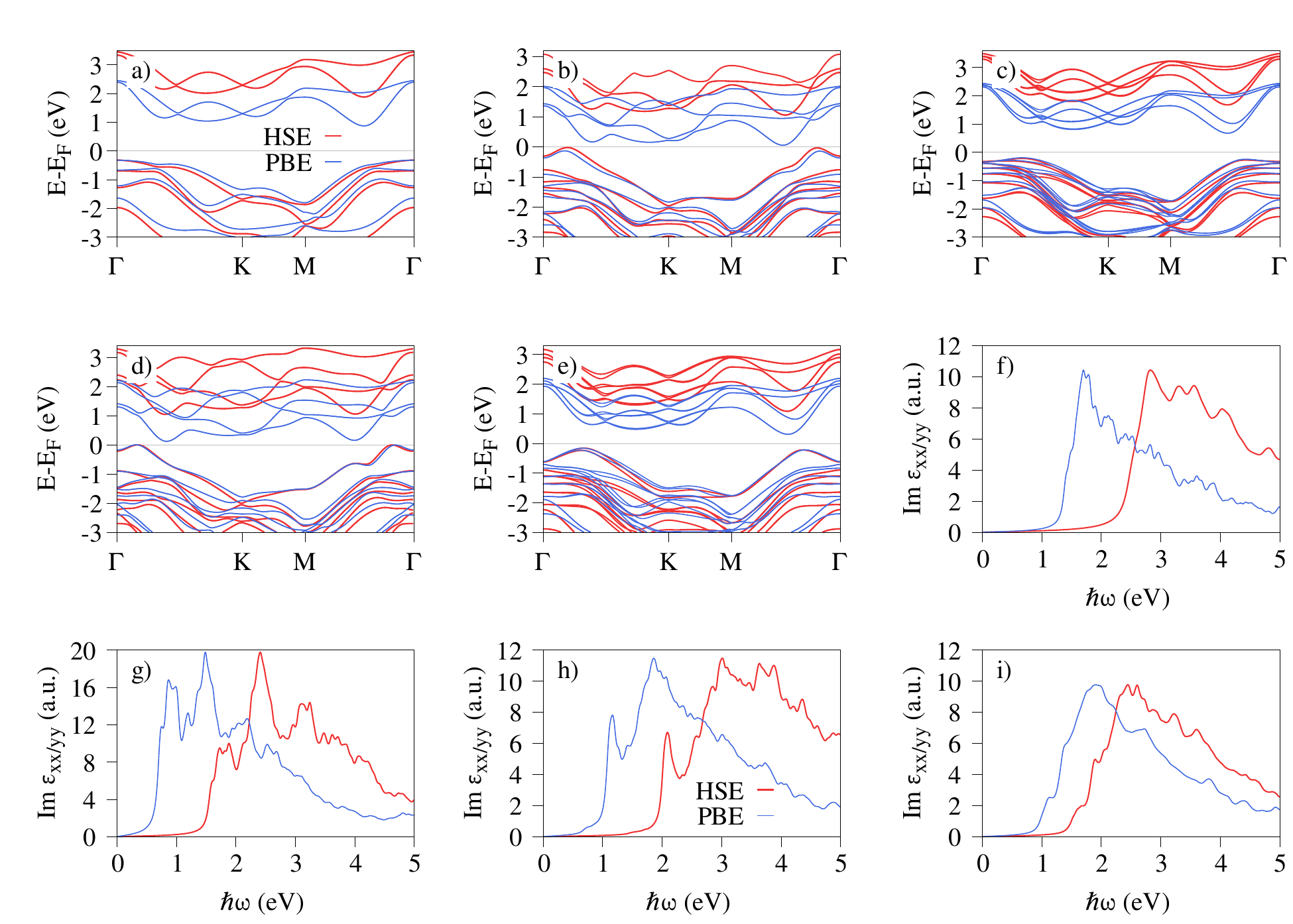}
    \caption{Comparison of PBE and hybrid HSE calculations. Relativistic band structures of the monolayer (a), AA bilayer ($\theta=0^\circ$) (b) and AB bilayer ($\theta=60^\circ$) PtSe$_2$. (d),(e) - band structures for AA and AB stackings with applied hydrostatic pressure $\delta$d=0.6\,\AA. The corresponding plots of the  imaginary part of dielectric function are shown in (g) and (h), respectively. (f),(i) - imaginary part of the dielectric tensor for the monolayer and $\theta=21.79^\circ$ twisted bilayer PtSe$_2$. }
    \label{fig_bnd_1}
\end{figure*}

\begin{figure*}[bh]
    \centering
\includegraphics[width=0.6\textwidth]{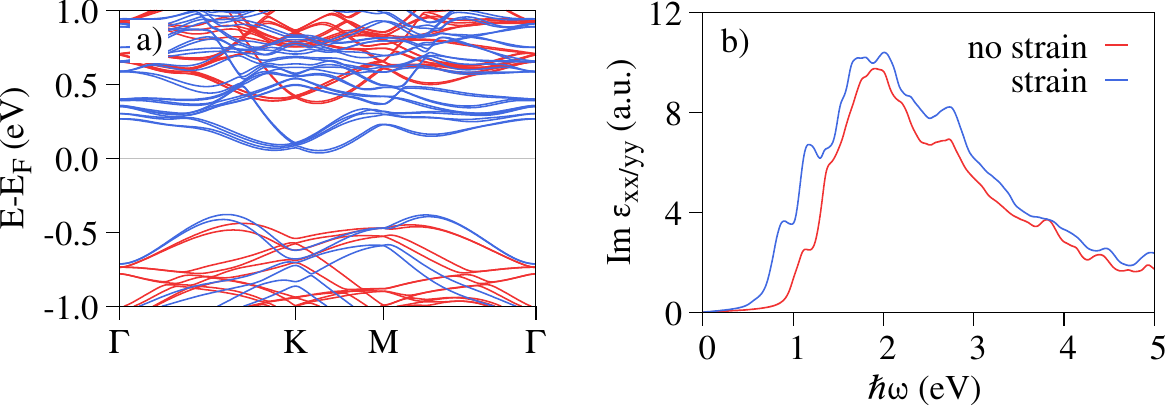}
    \caption{ Relativistic band structures (a) and dielectric function (b) for $\theta=21.79^\circ$- twist angle for unstrained (red line) and under hydrostatic pressure  $\delta d $ = 0.6\,\AA bilayer PtSe$_2$. }
    \label{fig_sup:21_06}
\end{figure*}

\end{document}